\documentclass[aps,prb,
preprint,
floatfix]{revtex4-2}

\usepackage{multirow}
\usepackage{amsmath}
\usepackage{mathtools}
\usepackage{amssymb}
\usepackage{mathrsfs}
\usepackage{graphics}
\usepackage{graphicx}
\usepackage{rotating}
\usepackage[table]{xcolor}
\usepackage{bbold}
\usepackage{bm}
\usepackage{bbm}
\usepackage{tikz}
\usetikzlibrary{arrows,positioning, shapes,intersections,calc}
\usepackage{hyperref}

\begin{document}

\title{Band representation, band connectivity and irreducibility}


\author{Jing Zhang}
\affiliation{Department of Physics and Institute of Advanced Studies, Tsinghua University, Beijing 100084, PRC \\ and Blackett Laboratory, Department of Physics, Imperial College London, Prince Consort Road, London SW7 2AZ, UK}

\date{\today}

\begin{abstract}
The ``band representation'', formulated by Zak in the 80's, is widely used in recent studies of topological phase of material. Elementary band representations (EBR) are considered the building block of topological quantum chemistry. However, there were debate on whether they admit split bands, or if they contain band invariant subspaces. This manuscript presents a clear definition of the basis and illustrate that band transformation properties described by EBR are dependent on the topology of band connectivity. An alternative derivation of analytical transformation properties of the band representation is given for the vector space spanned by an EBR. Two different Fourier transform conventions in defining the band representation induced from the same set of localised Wannier functions are discussed and identified as related by a simple gauge transformation. The tight binding band representations are used to derive the form of tight binding Hamiltonian using group theoretical technique. Their transformation properties allows explicit decomposition of EBR in terms of irreps of space group and use of projection operators in determining symmetry permitted band connectivity. It is shown that EBRs arising from Wykocff positions with multiplicity can have multiple atomic limits and identifies explicitly what occurs at high symmetry points on the surface of Brillouin zone as what determines the configuration of band connectivity and the irreducible nature of the connected band. Such multiple configurations arise from multiple ways of block-diagonalising representation matrix at these HSPs. The symmetry permitted connectivity also allows the existence of dynamic band invariant subspaces known as irreducible band representations (IBR). Such IBR, arising from Wannier functions centred on linear combination of orbits of Wyckoff positions, can also arise from interaction among composite EBRs. They are rooted in the permutation symmetry of crystallographic orbits of Wyckoff positions with multiplicity, and can occur even if the parental EBR is classified as indecomposable. When symmetry permitted split EBR occurs, which does not correspond to division between connected invariant subspaces, the topologies of the bands are potentially non-trivial on both sides of the gap. The work here provides the framework for symmetry constraint on Berry phases investigated in the accompanying manuscript.
\end{abstract}


\maketitle
\section{Introduction}
Band representation (BR), originally formulated by Zak\cite{Zak_J:1981A, Zak_J:1982A}, is receiving attention due to its application in studies of topology in the electronic band structure in wave-vector space\cite{Bradlyn_B:2017, Po_H_C:2017, Tang_F:2019A, Tang_F:2019B, Po_H_C:2020, Cano_J:2018B, Cano_J:2021, Vergniory_MG:2017}. To utilise BR in the study of topology of band structure, Elementary Band Representation (EBR) is considered playing a central role \cite{Bradlyn_B:2017, Po_H_C:2017,Cano_J:2018B}. The original work was extended by Evarestov and Smirnov\cite{Evarestov_R_A:1984A, Evarestov_R_A:1984B, Evarestov_RA:2007} and used in the tabulation of band representation (BANDREP) in the Bilbo Crystallgraphic Server\cite{Elcoro_L:2017}. Much of the topological quantum chemistry work is dubbed as the study of band structure without the Hamiltonian. There is an extensive description of group theoretical approach in condensed matter physics, including BR\cite{[{}][{ and articles in the same volume.}]Michel_L:2001A}. Central to its application is the band connectivity (continuity chords\cite{Zak_J:1982B}) between high symmetry points (HSP) in the Representation Domain (RD) of the first Brillouin Zone (BZ). The coincidence of band connectivity is one of the necessary conditions for equivalence between any two BRs. There are existing analysis of band connectivity \cite{Zak_J:1981A, Evarestov_RA:2007, Bradlyn_B:2017, Vergniory_MG:2017, Kruthoff_J:2017} but the work is largely restricted to decomposition of BR at HSPs, compatibility relations, and graph analysis/combinatronics. The basic underlying constraints are still compatibility relations\cite{Bouckaert_L_P:1936}. A more theoretical work on band connectivity was carried out by Barcy\cite{Barcy_H:1993}. There is an absence of clear methodology of analysing band connectivity and/or establishing connectivity under the atomic limit. The concept of irreducibility and equivalence of BRs were generally discussed as intrinsic properties of the basis without the context of connectivity.

The BRs are infinite dimensional representations of the space group. They are realised from Fourier transform of Wannier functions localised (centred) on the crystallographic orbits of equivalent Wyckoff positions. In defining the BR in k-space, there are two conventions which lead to the tight binding (TB) basis and the Zak basis. The original Slater-Koster formulation\cite{Slater_J:1954} of the TB theory is based on the TB basis whereas the recent analysis of band topology are based on the Zak. In the study of band topology, the concept of `Elementary Band Representation', `Composite Band Representation (CBR)' and `Physical Elementary Band Representation(PEBR)' were introduced\cite{Bradlyn_B:2017}. There had been claims\cite{Zak_J:1980, Michel_L:2001} and counter claims\cite{Cano_J:2018A,Po_H_C:2018} of full connectivity of EBR. These may be due to a lack of awareness of multiple ways to block-diagonalise the representation matrix at HSPs on the surface of BZ, and in turn the existence of multiple atomic limits for EBR arising from Wyckoff positions with multiplicity, and the existence of dynamic IBRs as invariant subspaces of EBR. These are the central theme of this manuscript.

The EBRs are described as in-decomposable/decomposable in BANDREP. The equivalence between any two BRs or CBR requires that the decomposition of the two BRs in-terms of irreducible representations (irrep) of space group $\mathcal{G}^{\rm\bf k}$ to be the same direct sum at all the points within the BZ. As shall be shown, there is the additional requirement of the same connectivity of the two sets of associated bands. Equivalence between two BRs guarantees the same decompositions of irreps of $\mathcal{G}$ occurs on all HSPs in the BZ for the connected bands. This is a necessary condition for the identification of equivalence between two BRs or CBRs. To examine topological phase transitions under symmetry compliant deformation of the Hamiltonian, it is also desirable to have a symmetry compliant TB Hamiltonian which includes all symmetry permitted interactions. Such model can be used in an analogous way to the Su-Schrieffer-Heager model for poly-acetylene for clarification of topological order and phase transitions (see Ref.\cite{Zhang_J:2024}). 

In this manuscript, the transformation properties of the full vector spaces spanned by the TB \& Zak basis are derived explicitly (Appx. \ref{sec:App2}) for a general space group ($\mathcal{G}$). A tight binding Hamiltonian is constructed using group theoretical method utilising the TB basis (Appx. \ref{sec:III}). The analytical form of the representation matrices of BR allows construction and use of group theoretical tools such as projection operators.  The transformation properties of the vector space spanned by EBR given in the literature is not complete due to existence of multiple ways to decompose invariant sub-space at HSPs on the BZ. In Sec.\ref{sec:sym_tools}, a general method is developed for the decomposition of EBR into irreducible representations (irreps) of $\mathcal{G}$ at ${\rm\bf k}$ using Frobenius reciprocity theorem. Projection operators are defined, providing tools working towards band connectivity. In Sec.\ref{sec:connectivity}, compatibility relations and projection operators are developed to establish links between nodes of EBR at different HSPs under the atomic limits. The concept of simple and complex links are introduced based on the outcome of successive application of projection operators at the two nodes/terminals at two distinct ${\rm\bf k}$ points (or HSPs). The resolution of complex links into unique links can lead to symmetry permitted disconnected bands. This disconnected configuration of connectivity can occur along decomposable EBRs. They show possible existence of multiple atomic limits in EBRs induced from Wannier functions centred on Wyckoff position with multiplicity. This methodology in establishing band connectivity goes beyond what can be obtained using compatibility relations and graph theory\cite{Vergniory_MG:2017}. In Sec.\ref{sec:irreducible_br}, the concepts of IBRs and equivalent BRs are defined. The IBRs are induced from Wannier functions centred on linear combination crystallographic orbits of Wyckoff positions with multiplicity. IBRs have their origin in real space symmetry and follows restoration of band irreducibility at HSPs on the surface of BZ by dynamic interaction. The choice of distinct atomic limits/band connectivity correspond to selection among possible ways of block-diagonalising the representation matrix at HSPs on the surface of BZ. Such dynamically occurring type 2 IBRs are equivalent to EBRs centred on Wyckoff position with no multiplicity. The existence of dynamic band invariant subspaces in EBR also means the concept of equivalence of BRs and discussion of band topology is not possible without the role of Hamiltonian (interaction). These concepts are illustrated in Sec.\ref{sec:connectivity} and \ref{sec:irreducible_br} with EBRs on a honeycomb lattice with sub-periodic layer group $\mathcal{G}=80({\rm L})$. The elucidation of concepts such as multiple atomic limit, IBRs and equivalence of BRs provides the framework of accompanying manuscript\cite{Zhang_J:2024} on symmetry constraints on Berry phase.

\section{Symmetry analysis of Elementary Band Representations}
\label{sec:sym_tools}
\subsection{Definition and transformation properties}
EBR is an infinite dimensional representation of the space group $\mathcal{G}$ induced from Wannier functions $W^{\mu, i}_{\bm{\tau}_\alpha}({\rm\bf r}-({\rm\bf t}_\ell+\tau_\alpha))$ centred on the crystallographic orbits of equivalent Wyckoff positions indexed by the primitive cell $\ell$ and equivalent Wyckoff position index $\alpha$. ${\rm\bf t}_\ell$ is a lattice translation and $\mu$ labels the transformation properties of $W$ as irrep $\mu$ of the site symmetry group ${\rm G}^\tau$. In real space, these infinite number of Wannier functions are closed under the action of the space group and form an infinite dimensional representations of $\mathcal{G}$.  

Taking a Fourier transform with respect the primitive cell location (${\rm\bf t}_\ell$) of the centre variable of these Wannier functions defines the EBR,
\begin{equation}
\Phi^{\rm\bf k}_{\bm{\tau}_\alpha, \mu, i}({\rm\bf r})= \left<{\rm\bf r}|{\rm\bf k},{\bm \tau}_\alpha,\mu,i\right>_{\rm Zak}=\Omega^{-1}\sum_{\ell}\exp\{i{\rm\bf k}\cdot {\rm\bf t}_\ell\}W^{\mu, i}_{\bm{\tau}_\alpha}({\rm\bf r}-({\rm\bf t}_\ell+\bm{\tau}_\alpha)).\label{eqn:Zak_basis}
\end{equation}
These are the original definition of Zak\cite{Zak_J:1981A}. An alternative definition takes the Fourier transform with respect to the localisation centre $({\rm\bf t}_\ell+\bm{\tau}_\alpha)$ and is used as basis of Slater Koster TB model\cite{Slater_J:1954}:
\begin{equation}
\phi^{\rm\bf k}_{\bm{\tau}_\alpha, \mu, i}({\rm\bf r})= \left<{\rm\bf r}|{\rm\bf k},{\bm \tau}_\alpha,\mu,i\right>_{\rm TB}=\Omega^{-1}\sum_{\ell}\exp\{i{\rm\bf k}\cdot ({\rm\bf t}_\ell+\bm{\tau}_\alpha)\}W^{\mu, i}_{\bm{\tau}_\alpha}({\rm\bf r}-({\rm\bf t}_\ell+\bm{\tau}_\alpha)).\label{eqn:TB_basis}
\end{equation}
The two are related by a ${\rm\bf k}$ dependent gauge transformation:
\begin{equation}
\phi^{\rm\bf k}_{\bm{\tau}_\alpha,\mu, i}({\rm\bf r})=\exp(i{\rm\bf k}\cdot\bm{\tau}_\alpha) \Phi^{\rm\bf k}_{\bm{\tau}_\alpha, \mu, i}({\rm\bf r}).
\label{eqn:gauge_transform}
\end{equation}
The dimension of the BR of $\mathcal{G}$ at a given wave vector ${\rm\bf k}\in$ RD is the product of multiplicity of the Wyckoff position, the dimension of irrep $\mu$ of the site symmetry group, and the number of arms in the star of ${\rm\bf k}$. If one considers ${\rm\bf k}\in$ BZ as linearly independent, then the dimensionality at a given wave vector is the product of multiplicity of the Wyckoff position, the dimension of irrep $\mu$ of the site symmetry group.

\begin{figure}[t]
\includegraphics[width=0.9\textwidth]{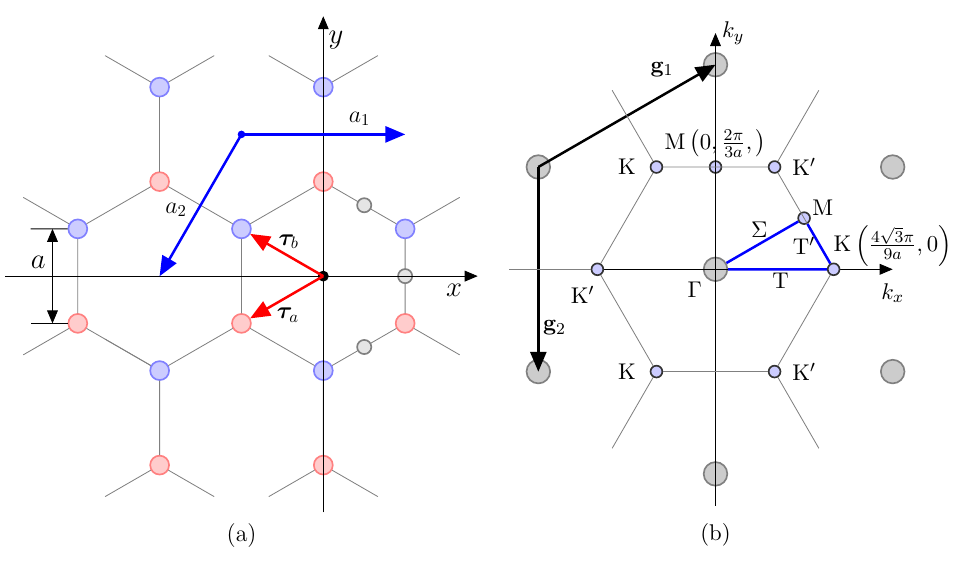}
\caption{(a) Honeycomb lattice ($\mathcal{G}=80({\rm L})$) indicating the position of Wyckoff position 2b ($\bigcirc$), 3d ($\circ$) and 1a($\bullet$) consistent with Ref.[\onlinecite{ITFCE}],  and (b) first Brillouin zone in reciprocal space and representation domain. The first BZ is the hexagon with vertices labelled K and K$^\prime$, and the RD is the $\bigtriangleup\Gamma{\rm MK}$. }
\label{fig:graphene}
\end{figure}

Fig.\ref{fig:graphene} illustrate the real and reciprocal space representation of the  honeycomb lattice with sub-periodic layer group $\mathcal{G}=80$(L). The RD is identified by $\bigtriangleup\Gamma{\rm MK}$. The Wyckoff positions 1a, 2b, and 3d are also indicated. The operations of element of $\tilde{\rm G}^{\rm\bf k}$ at different HSPs and site symmetry group ${\rm G}^\tau$ are identified in Tab.\ref{tab:G80-op}. 

An EBR may be expressed as a direct sum of representations of $\mathcal{G}$ in the RD\cite{Barcy_H:1993}:
\begin{equation}
\Gamma_{\rm G^{\tau}\uparrow}^\mathcal{G}
=\int_{\kappa\in {\rm Rep\ dom}}D^{\kappa}(g)\mathrm{d\kappa}\oplus D^\Gamma\oplus D^{\rm K}\oplus D^{\rm M}+\ldots
\label{eqn:EBR_direct_sum}
\end{equation}
where the integral part depends on the Born-von Karman group or the size of the crystal. The representation at HSPs such as $\Gamma$, K, M \ldots are sub representations independent of size of the crystal. At any given ${\rm\bf k}\in$ RD, these sub-representations are decomposable into irreps of $\mathcal{G}$. These irreps at different HSPs serves as nodes/terminals and are connected to one and another by irreps arising from decomposition of $D^\kappa(g){\rm d}\kappa$, giving rise to configuration of band connectivity in the BZ. The configuration of connectivity is not unique (see Sec.\ref{sec:connectivity}) and may split into disconnected bands\cite{Cano_J:2018A} (gapped state). They arise from {\em multiple-ways} a sub-representation at HSPs on the surface of BZ can be block diagonalised yielding the same decomposition\footnote{This statement means that there are multiple distinct unitary similarity transforms which will block diagonalise the representation matrices at these HSPs yielding the same direct sum of irreps of $\mathcal{G}$. This situation arises in EBR induced from Wannier functions centred on Wyckoff positions with multiplicity. They can also lead to different band topology in gapped systems.}. If a gapped state occurs, the whole EBR $\Gamma_{\rm G^{\tau}\uparrow}^\mathcal{G}$ may be decomposable into separate component {\em band representations} when the component band representation correspond to non-zero invariant subspaces of those described by the whole EBR for all ${\rm\bf k}\in$ BZ. Such component band representations are called IBRs. The {\em band irreducibility} here refers to equality of dimension of IBR and the connected band and spanning of the vector space of the band by the IBR basis at all ${\rm\bf k} \in$ BZ. When gapped system occurs which does not lead to IBRs, the connected bands on either side of the gap is symmetry indicated as potentially topologically non-trivial. Thus the transformation properties of band described by EBR are dependent on the band connectivity when all ${\rm\bf k}\in$ BZ are considered. The IBR turns out to be the building block of connected bands which are topologically trivial\cite{Zhang_J:2024}.

Space group is generally defined on the symmetry of infinite crystals. Bring in boundary conditions, even if they are periodic, may bring somewhat unexpected outcomes\cite{Barcy_H:1988}. It is vitally important to define what is meant by equivalent BRs (see Sec.\ref{sec:irreducible_br}) and derive impact of symmetry on the properties of such BRs based on basic principles.

The transformation properties of EBR given in the literature, for example Eq.(5) of Ref.\onlinecite{Cano_J:2018B}, generally makes no assumption on the topology of the band described by it. They describe the transformation properties of the vector space spanned by the whole EBR at any given ${\rm\bf k}\in$ RD, but contains no information on how such vector space may be decomposed into invariant sub-spaces (irreps of $\mathcal{G}$). This is particularly true if there are multiple ways to implement the decomposition at HSPs on the surface of BZ leading to split bands. i.e. they are the correct description of transformation properties of direct sum (over ${\rm\bf k}$) of vector spaces associated with it but contains no information on how such spaces may be joined together as bands. Therefore, these expressions are {\em incomplete description} of bands described by the EBR. However, projection operators based on these transformation properties allows identification of symmetry compliant band connectivity and in turn identification of IBRs and topologically trivial phases of connected bands\cite{Zhang_J:2024}. 

Distinct band connectivity arises from distinct similarity transformations which block diagonalise the representation matrices of EBR at HSPs on the surface of the BZ. The transformation properties of full vector space spanned by an EBR in the literature are not given in the form of representation matrices. Its equivalent is derived in Eq.\eqref{eqn:tfm} of Appx. \ref{sec:App2}. Eq.\eqref{eqn:tfm} can be obtained from Eq.(5) of Ref.\onlinecite{Cano_J:2018B} by recognising ${\rm\bf t}_{\beta\alpha}=h{\rm\bf q}_\alpha-{\rm\bf q}_\beta=R{\rm\bf q}_\alpha-{\rm\bf q}_\beta+{\rm\bf v}+{\rm\bf t}$ where $h=\{R|{\rm\bf v+t}\}$ is a general element of $\mathcal{G}$. Representation matrices are shown in Eq.\eqref{eqn:EBR_matrix} and Eq.\eqref{eqn:EBRTB_matrix} for the Zak and TB basis. The latter is particular useful due to the absence of explicit ${\rm\bf k}$ dependence on the rotational part. This not only facilitate the decomposition of EBR at any ${\rm\bf k}$ in terms of irreps of $\mathcal{G}$, but also enables path integral of Berry-Wilczek-Zee connections with respect to IBRs/EBRs to form representations of the space group. 

A central theme of this manuscript is the existence of dynamic band invariant subspace (IBR) in an EBR induced from Wannier functions centred on Wyckoff position with multiplicity. Associated with it, is the existence of multiple atomic limits for such EBRs and topology (connectivity) dependence of the transformation properties of band described by these basis at HSPs on the surface of the BZ. These are concepts not generally known in the literature, but have a profound impact on the study of symmetry constraint in the Berry phase\cite{Zhang_J:2024}.

\subsection{Decomposition of EBRs in terms of irreps of $\mathcal{G}$}
The energy eigenstates in the band theory are labelled by the irreps of $\mathcal{G}$ which are induced from irreps of $\mathcal{G}^{\rm\bf k}$ at ${\rm\bf k}$. At HSPs in the BZ, irreps from decomposition of EBR serve as nodes with terminals in the BZ and much of the work of quantum chemistry relates to establishing connectivity\cite{Zak_J:1982B} between these nodes/terminals(irreps) under the description of EBRs. The unitary irrep of $\mathcal{G}$ is normally induced from irrep of the little group of ${\rm\bf k}$: $\mathcal{G}^{\rm\bf k}$. These, in turn, are induced from and labelled by the irreps of the little co-group of ${\rm\bf k}$: $\tilde{{\rm G}}^{\rm\bf k}$. Through Frobenius reciprocity theorem, one obtains the desired decomposition of EBR at ${\rm\bf k}$ in terms of irreps of $\mathcal{G}$ by examining the the decomposition of the subduced representation of $\mathcal{G}^{\rm\bf k}$ from EBR in terms of irreps of $\mathcal{G}^{\rm\bf k}$. Much of the theory on induced representation of space group is covered in chapter 4 of Ref.\cite{Bradley_C_J:2010}. 

Symmetry investigation of band structure often lead to the ``sub-group method''\cite{Lax_M:1965,Birman_J:1966}. Where such method is applicable, one expect any summation over the translation sub-group to be factored out in application of the group rearrangement theorem\cite{Lax_M:2001}, leaving only sums over rotational part of the symmetry operation (coset representatives in the decomposition of $\mathcal{G}$ in Eq.\eqref{eqn:A1}). Symmetry analysis of EBR is no exception. As the Zak and TB basis differ only by a ${\rm\bf k}$ dependent gauge term, they span the same vector spaces and have the same decomposition in terms of irreps of $\mathcal{G}$ for all ${\rm\bf k}\in$ BZ. The transformation property of EBR shown in Eq.\eqref{eqn:EBRTB_matrix} of Appx.\ref{sec:App2} is better suited for this purpose. The absence of explicit ${\rm\bf k}$ dependence in the rotational part of Eq.\eqref{eqn:EBRTB_matrix} for the TB basis makes it significantly easier to apply symmetry tools such as decomposition in terms of irreps and projection operators. Without loss of generality, the work follows is based on the TB basis. 

The subduced representation of $\mathcal{G}^{\rm\bf k}$ from a fully connected EBR is obtained from Eq.\eqref{eqn:EBRTB_matrix} by setting ${\rm\bf k}^\prime={\rm\bf k}$ and focusing on the relevant diagonal block indexed by ${\rm\bf k}\in$ RD:
\begin{eqnarray}
\Gamma_{\mathcal{G}(\tilde{\rm G}^{\bm{\tau}}_\mu\uparrow)\downarrow}^{\mathcal{G}^{\rm\bf k}}(\{R|{\rm\bf v+t}\})_{{\bm{\tau}_{\alpha^\prime}, i^\prime };\,{\bm{\tau}_\alpha}, i} &=&\exp\{-iR{\rm\bf k}\cdot({\rm\bf v+t})\} \exp \{i{\rm\bf g}_R\cdot\bm{\tau}_{\alpha^\prime}\} D^{\bm{\tau},\mu}(R)_{\bm{\tau}_{\alpha^\prime} i^\prime,  \bm{\tau}_\alpha i}\notag \\
&=&\exp\{-i{\rm\bf k}\cdot({\rm\bf v+t})\} \exp \{-i{\rm\bf g}_R\cdot {\rm\bf v}\}\exp \{i{\rm\bf g}_R\cdot\bm{\tau}_{\alpha^\prime}\} D^{\bm{\tau},\mu}(R)_{\bm{\tau}_{\alpha^\prime} i^\prime,  \bm{\tau}_\alpha i}\notag \\
&&~~~~\{R|{\rm\bf v+t}\}\in\mathcal{G}^{\rm\bf k},\;\;R{\rm\bf k}\equiv {\rm\bf k}. \label{eqn:subTB_BR}
\end{eqnarray}
Here the term $\exp \{i{\rm\bf g}_R\cdot\bm{\tau}_{\alpha^\prime}\}$ is known as the gauge term. It can only takes non-unity value at HSPs on the surface of BZ where ${\rm\bf g}_R\ne{\bf 0}$.

The irrep of $\mathcal{G}^{\rm\bf k}$ is induced from irrep (or projective irrep) of its co-group $\tilde{\rm G}^{\rm\bf k}$. Its matrix of representation of $\mathcal{G}^{\rm\bf k}$ induced from irrep $p$ of $\tilde{\rm G}^{\rm\bf k}$ is often given by:\cite{Bradley_C_J:2010}
\begin{equation}
\Gamma^{\mathcal{G}^{\rm\bf k}}_{\tilde{\rm G}^{\rm\bf k}_p\uparrow}(\{R|{\rm\bf v+t}\})	=\exp\{-i{\rm\bf k}\cdot{\rm\bf (v+t)}\}D_p^{\rm\bf k}(R), \;\;\;\;\{R|{\rm\bf v+t}\}\in \mathcal{G}^{\rm\bf k}\label{eqn:Co_Gofk}
\end{equation}
where $p$ labels the vector irrep of $\tilde{\rm G}^{\rm\bf k}$ at any ${\rm\bf k}\in$ BZ or the projective irrep of $\tilde{\rm G}^{\rm\bf k}$ at HSPs on the surface of BZ of non-symmorphic space group. These are used to induce irreps of space group $\mathcal{G}$. The irreducibility of such induced representation is determined by Mackey's or Johnston's irreducibility criterion\cite{Bradley_C_J:2010}. Implicit in this process is the requirement of the gauge condition in Eq.\eqref{eqn:Zak_gauge} except where projective representation occur in the case of non-symmorphic space group. 

The choice of Eq.\eqref{eqn:Co_Gofk} is not very convenient for intended decomposition of BRs because of the differing dependence of ${\rm\bf k}$ and $R{\rm\bf k}$ in the exponent containing ${\rm\bf v+t}$. There is an alternative choice for the representation matrices as
\begin{eqnarray}
\Gamma^{\mathcal{G}^{\rm\bf k}}_{\tilde{\rm G}^{\rm\bf k}_p\uparrow}(\{R|{\rm\bf v+t}\})	&=&\exp\{-i{\rm\bf k}\cdot{\rm\bf (v+t)}\}\exp\{-i{\rm\bf g}_R\cdot{\rm\bf v}\}D_p^{\rm\bf k}(R), \notag \\
&=&\exp\{-iR{\rm\bf k}\cdot{\rm\bf (v+t)}\}D_p^{\rm\bf k}(R), \;\;\;\;\{R|{\rm\bf v+t}\}\in \mathcal{G}^{\rm\bf k}\;\;\;\;R{\rm\bf k}\equiv {\rm\bf k}.\label{eqn:Co_Gofka}
\end{eqnarray}
The situation where $R{\rm\bf k}$ and ${\rm\bf k}$ are linearly dependent can only occur at HSPs on the surface of the BZ where $R{\rm\bf k}={\rm\bf k}+{\rm\bf g}_R$ ($\tilde{\rm\bf g}_R=R^{-1}{\rm\bf k}-{\rm\bf k}=-R^{-1}{\rm\bf g}_R$). One can see the equivalence of the two through the following. The difference between the two expression is given by $\exp\{i{\rm\bf g}_R\cdot({\rm\bf v+t})\}$ where $R{\rm\bf k}-{\rm\bf k}={\rm\bf g}_R\ne 0$. This can only occur on HSP on the surface of BZ. For symmorphic space groups, ${\rm\bf v}=\bm{0}$ and the term returns 1 irrespective the value of ${\rm\bf g}_R$ (reciprocal lattice vector). For non-symmorphic space group and where ${\rm\bf g}_R=0$, the term also returns 1. The only remaining case is non-symmorphic space group at HSP on the surface of BZ where projective representation is required. Specifically Eq.\eqref{eqn:Co_Gofk} yields factoring system
\[
D^{\rm\bf k}_p(R_i)D^{\rm\bf k}_p(R_j)=\exp\{i(\tilde{\rm\bf g}_i\cdot R_i{\rm\bf v}_j)\}D^{\rm\bf k}_p(R_k)\quad\quad\quad\mbox{where~}R_i^{-1}{\rm\bf k}-{\rm\bf k}=\tilde{\rm\bf g}_i=-R_i^{-1}{\rm\bf g}_i
\] 
where as Eq.\eqref{eqn:Co_Gofka} yields
\[
D^{\rm\bf k}_p(R_i)D^{\rm\bf k}_p(R_j)=\exp\{i(\tilde{\rm\bf g}_i\cdot R_i{\rm\bf v}_j)\}\exp\{i({\rm\bf g}_i\cdot{\rm\bf v_i}+{\rm\bf g}_j\cdot{\rm\bf v_j}-{\rm\bf g}_k\cdot{\rm\bf v_k}\}D^{\rm\bf k}_p(R_k).
\] 
Thus the difference between the two choice is merely a difference in factoring system when projective representation is required at HSPs on the surface of BZ under non-symmorphic space group. The additional term in the factoring system is associated with each element and does not change the associativity of the existing factor system. We wish to examine the decomposition of the subduced band representation at given ${\rm\bf k}$ into irreducible form upon restriction to $\mathcal{G}^{\rm\bf k}$. For this purpose, it is easier to make the choice of Eq.\eqref{eqn:Co_Gofka} as irreducible representation of $\mathcal{G}^{\rm\bf k}$. This will allow the irreps of $\mathcal{G}^{\rm\bf k}$ to share the same dependence on translation as the BR, and factoring out many of the summation over $\mathcal{T}$ we encounter. It will also lead to the same factoring system between the subduced EBR and irrep of $\mathcal{G}^{\rm\bf k}$ in the case of non-symmorphic space group at HSPs on the surface of BZ. The character of the irrep of $\mathcal{G}^{\rm\bf k}$ shown in Eq.\eqref{eqn:Co_Gofka} is:
\begin{equation}
\chi^{\mathcal{G}^{\rm\bf k}}_{\tilde{\rm G}^{\rm\bf k}_p\uparrow}(\{R|{\rm\bf v+t}\})=\exp\{-iR{\rm\bf k}\cdot{\rm\bf (v+t)}\}\chi^p_{\tilde{\rm G}^{\rm\bf k}}(R), \quad\quad\{R|{\rm\bf v+t}\}\in \mathcal{G}^{\rm\bf k}\label{GoK_char}
\end{equation}
From Eq.\eqref{eqn:subTB_BR}, the character of the subduced TB EBR, on restriction to ${\rm\bf k}$ and $\mathcal{G}^{\rm\bf k}$ is given by:
\begin{align}
&\chi^{\mathcal{G}^{\rm\bf k}}_{\mathcal{G}(\tilde{\rm G}^{\bm{\tau}}_\mu\uparrow)\downarrow}(\{R|{\rm\bf v+t}\})=\exp\{-R{\rm\bf k}\cdot{\rm\bf (v+t)}\}\notag \\
&\quad\quad\quad\quad\sum_{\bm{\tau}_\alpha}\delta_{\bm{\tau}_\alpha, \{R|{\rm\bf v+t}\}\bm{\tau}_\alpha}\exp \{i{\rm\bf g}_R\cdot\bm{\tau}_{\alpha}\}\chi^\mu_{\tilde{\rm G}^{\bm{\tau}}}({R^{\bm{\tau}}_\alpha}^{-1}RR^{\bm{\tau}}_\alpha) \quad\quad\{R|{\rm\bf v+t}\}\in \mathcal{G}^{\rm\bf k}\label{BR_char}
\end{align}
where $R{\rm\bf k}-{\rm\bf k}={\rm\bf g}_R$ can take the value of $\bm{0}$ or some reciprocal lattice vector (for some HSP on the surface of BZ and certain $R\in\tilde{\rm G}^{\rm\bf k}$)
The decomposition theorem gives the multiplicity of irrep $\Gamma^{\mathcal{G}^{\rm\bf k}}_{\tilde{\rm G}^{\rm\bf k}_p\uparrow} $ contained in $\Gamma^{\mathcal{G}^{\rm\bf k}}_{\mathcal{G}(\tilde{\rm G}^{\bm{\tau}}_\mu\uparrow)\downarrow} $  
\begin{eqnarray*}                    
a&=&\frac{1}{|\mathcal{G}^{\rm\bf k}|}\sum_{\{R|{\rm\bf v+t}\}\in \mathcal{G}^{\rm\bf k}} {\chi^{\mathcal{G}^{\rm\bf k}}_{\tilde{\rm G}^{\rm\bf k}_p\uparrow}(\{R|{\rm\bf v+t}\})}^\ast\chi^{\mathcal{G}^{\rm\bf k}}_{\mathcal{G}(\tilde{\rm G}^{\bm{\tau}}_\mu\uparrow)\downarrow}(\{R|{\rm\bf v+t}\})\\
&=& \frac{1}{|\mathcal{G}^{\rm\bf k}|}\sum_{\{R|{\rm\bf v+t}\}\in \mathcal{G}^{\rm\bf k}}\chi^p_{\tilde{\rm G}^{\rm\bf k}}(R)^\ast \sum_{\bm{\tau}_\alpha}\delta_{\bm{\tau}_\alpha, \{R|{\rm\bf v+t}\}\bm{\tau}_\alpha}\exp \{i{\rm\bf g}_R\cdot \bm{\tau}_{\alpha}\}\chi^\mu_{\tilde{\rm G}^{\bm{\tau}}}({R^{\bm{\tau}}_\alpha}^{-1}RR^{\bm{\tau}}_\alpha)
\end{eqnarray*}
The summation over the translation sub-group is factored out, giving
\begin{equation}
a=\frac{1}{|\tilde{\rm G}^{\rm\bf k}|}\sum_{R\in \tilde{\rm G}^{\rm\bf k}}\chi^p_{\tilde{\rm G}^{\rm\bf k}}(R)^\ast \underbrace{\sum_{\bm{\tau}_\alpha}\delta_{\bm{\tau}_\alpha, \{R|{\rm\bf v}\}\bm{\tau}_\alpha}\exp \{i{\rm\bf g}_R\cdot \bm{\tau}_{\alpha}\}\chi^\mu_{\tilde{\rm G}^{\bm{\tau}}}({R^{\bm{\tau}}_\alpha}^{-1}RR^{\bm{\tau}}_\alpha)}_{\chi_\phi(R)},\quad R\in\tilde{{\rm G}}^{\rm\bf k}.\label{eqn:band_decomp}
\end{equation}
For a given element of $R\in (\tilde{\rm G}^{\rm\bf k}\bigcap{\rm G}^{\bm{\tau},\alpha})$, its mapping to ${R^{\bm{\tau}}_\alpha}^{-1}RR^{\bm{\tau}}_\alpha\in \tilde{\rm G}^{\bm{\tau}}$ is well defined. Otherwise, the mapping to ${R^{\bm{\tau}}_\alpha}^{-1}RR^{\bm{\tau}}_\alpha\notin \tilde{\rm G}^{\bm{\tau}}$. Under such conditions, the term $\delta_{\bm{\tau}_\alpha, \{R|{\rm\bf v}\}\bm{\tau}_\alpha}$ yield zero and renders the mapping unnecessary. The decomposition of EBR at ${\rm\bf k}\in {\rm RD}$ in terms of irreps of $\mathcal{G}$ is then obtained through Frobenius reciprocity theorem.

With the exception of HSPs on the surface of the BZ where $R{\rm\bf k}-{\rm\bf k}={\rm\bf g}_R\ne\bm{0}$, the decomposition of an EBR is fixed everywhere in the Brillouin zone once the decomposition at $\Gamma$ point is specified. Thus to establish the equivalence of band representations, we need to establish {\em the same decomposition at $\Gamma$ point as well as all other high symmetry points on the surface of the Brillouin zone where $R{\rm\bf k}-{\rm\bf k}={\rm\bf g}_R\ne\bm{0}$ occurs}. The decomposition rule here only depends on the elements of $\tilde{\rm G}^{\rm\bf k}$ and the Wyckoff position with the summation over the translation subgroup factored out. It does not depend on lattice translation contained in $\tilde{{\rm G}}^{\bm{\tau},\alpha}$ or non-lattice translation ${\rm\bf v}$ contained in $\{R|{\rm\bf v+t}\}$.

A decomposition analysis is performed in Appx. \ref{sec:App3} for various EBRs centred on Wyckoff position 2b, 3d, and 1a of a honeycomb lattice. The operations of the site symmetry groups $\tilde{\rm G}^{\bm{\tau}}$ and little co-group of ${\rm\bf k}$ $\tilde{\rm G}^{\rm\bf k}$ are identified in Tab.\ref{tab:G80-op}. The value of the character $\chi_\phi$ of some EBRs are listed in Tab.\ref{tab:G80}. Where the gauge term modifies the character on HSPs on the surface of BZ are also indicated. The resulting decompositions of EBRs at HSPs are shown in Tab.\ref{tab:spG80_decomp}.

In evaluating the character of BR upon restriction to $\mathcal{G}^{\rm\bf k}$, the TB \& Zak basis return the same value. In both cases, the term $\exp[i{\rm\bf g}_R\cdot\bm{\tau}_\alpha]$ is significant at HSPs where ${\rm\bf g}_R\ne \bm{0}$. 

\subsection{Projection operator at ${\rm\bf k}$}
With the subduced representation of $\mathcal{G}^{\rm\bf k}$ from EBR matrices in Eq.\eqref{eqn:subTB_BR} and irrep of $\mathcal{G}^{\rm\bf k}$ in Eq.\eqref{eqn:Co_Gofka}, it is possible to define projection operators which would yield the linear combination of EBRs which transform as irrep of $\mathcal{G}^{\rm\bf k}$.
\begin{eqnarray}
\hat{\mathcal{P}}^{{\rm\bf k}, p}_{i,j}&=&\frac{1}{|\mathcal{G}^{\rm\bf k}|}\sum_{g\in\mathcal{G}^{\bf k}}{\Gamma^{\mathcal{G}^{\bf k}}_{\tilde{\rm G}^{\rm\bf k}_p\uparrow}(g)_{i,j}}^{\ast}\Gamma_{\mathcal{G}(\tilde{\rm G}^{\bm{\tau}}_\mu\uparrow)\downarrow}^{\mathcal{G}^{\rm\bf k}}(g)_{{\bm{\tau}_{\alpha^\prime}, i^\prime };\,{\bm{\tau}_\alpha}, j^\prime} \notag
\end{eqnarray}
Here ${\rm\bf k} \in $ BZ and $p$ is the irrep label of $\tilde{{\rm G}}^{\rm\bf k}$, $i$ and $j$ are the indices of irrep $p$, $i^\prime$ and $j^\prime$ are the indices of irrep $\mu$. The translation dependent terms of the two matrices cancel each other and the summation over translations is factored out. Thus one only need to sum over the elements of the little co-group of ${\rm\bf k}$.
\begin{eqnarray}
\hat{\mathcal{P}}^{{\rm\bf k}, p}_{i,j}&=&\frac{1}{|\tilde{\rm G}^{\rm\bf k}|}\sum_{R\in\tilde{\rm G}^{\bf k}}D^p(R)_{i,j}^{\ast}\exp\{i{\rm\bf g}_R\cdot\bm{\tau}_\alpha^\prime}\}D^{\tau,\mu}(R)_{{\bm{\tau}_{\alpha^\prime}, m^\prime ;\,{\bm{\tau}_\alpha}, m} 
\label{eqn:projection}
\end{eqnarray}
where ${\rm\bf g}_R$ is defined by Eq.\eqref{eqn:g_R}, and $D^{\tau,\mu}(R)$ is defined by Eq.\eqref{eqn:atomic_rep}.

\section{Band connectivity of EBR and Multiple Atomic Limits}
\label{sec:connectivity}
The symmetry analysis of EBR in the previous section established the decomposition of EBR in terms of unitary irrep of $\mathcal{G}^{\rm\bf k}$ at a given ${\rm\bf k}$. By Wigner Theorem, these irreps represent the transformation properties of energy eigenstate at ${\rm\bf k}$ under band theory. They serve as labels to nodes/terminals in the BZ (labelled by ${\rm\bf k}$, irrep label of $\mathcal{G}^{\rm\bf k}$ and its component index). Topological quantum chemistry involves the investigation of how these nodes/terminals are connected by the bands in the BZ, and the topological invariant associated with the band structure represented by the BR basis under certain hopping parameters. Behind the scene, the goal is really to find structure of invariant subspaces which are embedded with the connectivity. 

The compatibility relations\cite{Bouckaert_L_P:1936} ensures that irreps of nodes/terminals of immediate lower symmetry neighbours connected to HSPs are determined by those at HSPs. We can restrict the discussion to those at $\Gamma$ and HSPs on the surface of BZ. Much of the literature had been focused on the compatibility relation and graph theory. The major hurdle in determining the band connectivity of EBR in the atomic limit is to link HSPs which are {\em not} immediate neighbours to each other. The projection operator in Eq.\eqref{eqn:projection} is used to overcome this hurdle.

The tools available to identify EBR connectivity are: 1) decomposition of BR at HSPs assuming absence of accidental degeneracy, 2) compatibility relation/graph theory\cite{Vergniory_MG:2017}, and 3) projection operators one is able to construct to establish connectivity between nodes/terminals at HSPs which are {\em not} immediate neighbours.  We will see the methodology of compatibility relation/graph theory is imbedded within and improved upon by the projection operator technique developed here. In particular, the concept of complex link and dynamic irreducible BRs are introduced. Section \ref{sec:sym_tools} and Appx. \ref{sec:App3} describe the decomposition of an isolated EBR, limiting the possible irreps of $\mathcal{G}$ occurring at any wave vector ${\rm\bf k}\in {\rm RD}$. The following address the use of two remaining tools.

\subsection{Compatibility relations}
\label{sec:V1}
Compatibility relations links the symmetry of an eigenstate at HSP to that of its immediate neighbours of lower symmetry. This is the consequence of the group/subgroup relations that exist between the little co-group of ${\rm\bf k}$ ($\tilde{\rm G}^{\rm\bf k}$) at HSP and its immediate lower symmetry {\em neighbours}. The representation of the $\mathcal{G}^{\rm\bf k}$ at lower symmetry neighbours is obtained from subduction of representation at HSPs upon restriction to the $\mathcal{G}^{\rm\bf k}$ at lower symmetry neighbours. This has no direct link to band representations but they must be satisfied by a physical band and in an EBR. This helps identification of {\em local} band connectivity in EBRs. 

\begin{table}
\caption{\label{tbl:G80_comp}Compatibility relations between irreducible representations of $\tilde{\rm G}^{\rm\bf k}$ at high symmetry points $\Gamma$, K, M and its lower neighbours under $\mathcal{G}=80(L)$ given the relations of symmetry operations of $\tilde{\rm G}^{\rm\bf k}$ shown in Tab.\ref{tab:G80-op}.}
\begin{tabular}{|c|c|c|c|c|c|c|c|c|c|c|c|c|}\hline
${\rm T}({\rm C}_{2v})\downarrow$ & ${\rm T}_1$ & ${\rm T}_3$ & ${\rm T}_3 \oplus {\rm T}_4$ & ${\rm T}_1 \oplus {\rm T}_2$ & ${\rm T}_4$ & ${\rm T}_2$ & ${\rm T}_1 \oplus {\rm T}_2$ & ${\rm T}_3 \oplus {\rm T}_4$ & ${\rm T}_5$ & ${\rm T}_5$ & ${\rm T}_5$\\ \hline
$\Gamma({\rm D}_{6h}$) & $\Gamma_1^+$ & $\Gamma_4^+$ & $\Gamma_5^+$ & $\Gamma_6^+$ & $\Gamma_2^-$ & $\Gamma_3^-$ & $\Gamma_5^-$ & $\Gamma_6^-$ & $\Gamma_7^\pm$ & $\Gamma_8^\pm$ & $\Gamma_9^\pm$ \\ \hline
$\Sigma({\rm C}_{2v})\downarrow$ & $\Sigma_1$ & $\Sigma_2$ & $\Sigma_2\oplus\Sigma_3$ & $\Sigma_1\oplus\Sigma_4$ & $\Sigma_2$ & $\Sigma_1$	 & $\Sigma_1\oplus\Sigma_4$  & $\Sigma_2\oplus\Sigma_3$ & $\Sigma_5$ & $\Sigma_5$ & $\Sigma_5$ \\ \hline 
\end{tabular}
\\
\begin{tabular}{|c|c|c|c|c|c|c|c|c|c|}\hline
${\rm T}\& {\rm T}^\prime({\rm C}_{2v})\downarrow$ & ${\rm T}_1$ & ${\rm T}_2$ & ${\rm T}_3$ & ${\rm T}_4$ & ${\rm T}_3 \oplus {\rm T}_4$ & ${\rm T}_1 \oplus {\rm T}_2$ & ${\rm T}_5$ & ${\rm T}_5$ & ${\rm T}_5$ \\ \hline
${\rm K}({\rm D}_{3h}$) & ${\rm K}_1$ & ${\rm K}_2$ & ${\rm K}_3$ & ${\rm K}_4$ & ${\rm K}_5$ & ${\rm K}_6$ & ${\rm K}_7$ & ${\rm K}_8$ & ${\rm K}_9$  \\ \hline
\end{tabular}
\\
\begin{tabular}{|c|c|c|c|c|c|c|c|c|c|c|}\hline
${\rm T}^\prime({\rm C}_{2v})\downarrow$ & ${\rm T}_1$ & ${\rm T}_4$ & ${\rm T}_2$ & ${\rm T}_3$ & ${\rm T}_5$ & ${\rm T}_3$ & ${\rm T}_2$ & ${\rm T}_4 $ & ${\rm T}_1$ & ${\rm T}_5$ \\ \hline
${\rm M}({\rm D}_{2h}$) & ${\rm M}_1^+$ & ${\rm M}_2^+$ & ${\rm M}_3^+$ & ${\rm M}_4^+$ & ${\rm M}_5^+$ & ${\rm M}_1^-$ & ${\rm M}_2^-$ & ${\rm M}_3^-$ & ${\rm M}_4^-$ & ${\rm M}_5^-$  \\ \hline
$\Sigma({\rm C}_{2v})\downarrow$ & $\Sigma_1$ & $\Sigma_3$ & $\Sigma_4$ & $\Sigma_2$ & $\Sigma_5$ & $\Sigma_3$	 & $\Sigma_1$  & $\Sigma_2$ & $\Sigma_4$ & $\Sigma_5$  \\ \hline 
\end{tabular}
\end{table}
\begin{figure}[b]
\includegraphics[width=0.9\textwidth]{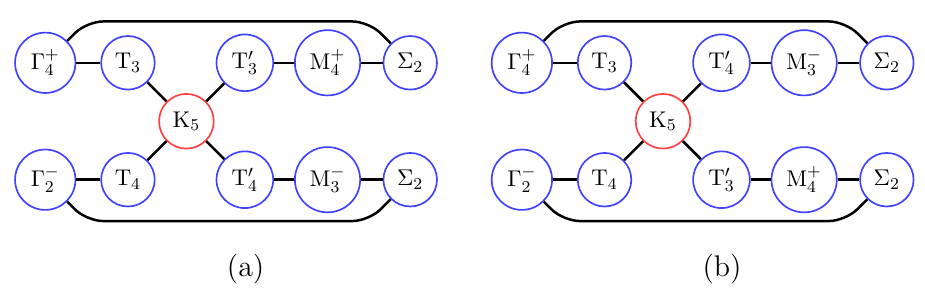}
\caption{Connected graph relating to the band structure, described by the spinless EBR induced from p$_z$ orbital centred on Wyckoff position 2b of a honeycomb lattice, based on compatibility relations and graph theory. The degeneracy that existed at K ensures the system is in semi-metal phase for half filling of the band. Only configuration (a) is symmetry permitted when it is determined from projection operators.}
\label{fig:connected}
\end{figure}
The characters of the irreps of $\tilde{\rm G}^{\rm\bf k}$  are readily available and the compatibility relation of irreducible representation of $\tilde{\rm G}^{\rm\bf k}$ at HSPs, such as $\Gamma$ with its lower symmetry neighbours can be established using the relations between $\tilde{\rm G}^{\rm\bf k}$ and ${\rm G}^\tau$ shown in Tab.\ref{tab:G80-op} and relevant character table of point groups\cite{Koster_G:1963}. For all the relevant irreps that occur in the decompositions at $\Gamma$, T, K, T$^\prime$, M and $\Sigma$, the compatibility relations are listed in Tab.\ref{tbl:G80_comp}. It is important to recognise that such compatibility relation is only {\em deterministic} from HSPs to its lower symmetry neighbours. No relation between the $\Gamma$, ${\rm K}$ or M points is uniquely determined by the compatibility relations as they are separated in ${\rm\bf k}$-space by T, ${\rm T}^\prime$ and $\Sigma$.  

It is possible to discuss symmetry allowed band connectivity using combination of compatibility relations, decomposition of BRs and graph theory\cite{Bradlyn_B:2017}. For example, the EBR induced from ${\rm p}_z$ orbital without spin centred on Wyckoff position 2b can generate connected HSPs as shown in Fig.\ref{fig:connected}. Only one connected component exists. There are two possible configurations permitted by the combination of compatibility relations, decomposition of BR and graph theory. The ambiguity between the two configurations remain under compatibility considerations. Both lead to semi-metal phase when the band is half filled. The use of analytical form of BRs and projection operators allows this ambiguity to be resolved with the configuration shown in Fig.\ref{fig:connected} (a) as the only symmetry permitted band connectivity (see end of next subsection) from the EBR. The configuration in Fig.\ref{fig:connected} (b) is symmetry indicated as topologically non-trivial.

\subsection{Connectivity in EBR and its atomic limits}
\label{sec:m_atomic_limit}
The method of using compatibility relations applies to relations between HSPs and its immediate neighbours. One wishes to develop relations between nodes/terminals on non-adjacent HSPs at ${\rm\bf k_1}$ and ${\rm\bf k_2}$, in the atomic limit. Atomic limit here refers to scaling the lattice parameter to infinity, or alternatively, letting the hoping parameters approach zero. Thus there is no ${\rm\bf k}$ dependent interaction/dispersion under the TB model. The only tools one has are symmetry based such as projection operators defined in Eq.\eqref{eqn:projection}. Suppose that there is some linear combination of an EBRs $\left|\eta\right>$ (a column vector) which yield a non-zero result in the following
\[
\mathcal{P}^{{\rm\bf k}_1,p}_{i,i}\mathcal{P}^{{\rm\bf k}_2,q}_{j,j}\left|\eta\right>\ne {\mathbb 0}.
\]
Then, in absence of any interaction/dispersion, the vector $\left|\xi\right>=\mathcal{P}^{{\rm\bf k}_2,q}_{j,j}\left|\eta\right>$ would evolve continuously from ${\rm\bf k}_2$ into some state $\left|\xi^\prime\right>$ at ${\rm\bf k}_1$. Since $\mathcal{P}^{{\rm\bf k}_1,p}_{i,i}\left|\xi^\prime\right>\ne {\mathbb 0}$, a link between node/terminal $^{{\rm\bf k}_2, q}_{j,j}$ and $^{{\rm\bf k}_1, p}_{i,i}$ must be possible. In fact, there are three possibilities
\begin{subequations}
\begin{eqnarray}
\mathcal{P}^{{\rm\bf k}_1,p}_{i,i}\mathcal{P}^{{\rm\bf k}_2,q}_{j,j}&=&\mathcal{P}^{{\rm\bf k}_2,q}_{j,j} \label{eqn:simple_lk} \\
\mathcal{P}^{{\rm\bf k}_1,p}_{i,i}\mathcal{P}^{{\rm\bf k}_2,q}_{j,j}&\ne& {\mathbb 0} \label{eqn:complex_lk} \\
\mathcal{P}^{{\rm\bf k}_1,p}_{i,i}\mathcal{P}^{{\rm\bf k}_2,q}_{j,j}&=& {\mathbb 0} \label{eqn:no_lk} 
\end{eqnarray}
\end{subequations} 
These three conditions correspond to Eq.\eqref{eqn:simple_lk}: an exclusive link between the two nodes/terminals called {\em simple link}; Eq.\eqref{eqn:complex_lk}: a possible link between the two nodes/terminals called {\em complex link}; and Eq.\eqref{eqn:no_lk}: no link can exist between the two nodes/terminals as next nearest neighbours, {\em under the atomic limit}. The complex link implies possible link may exist between multiple terminals where $\mathcal{P}^{{\rm\bf k}_1,p}_{i,i}\mathcal{P}^{{\rm\bf k}_2,q}_{j,j}= {\mathbb 0}$, $\mathcal{P}^{{\rm\bf k}_1,p^\prime}_{i^\prime,i^\prime}\mathcal{P}^{{\rm\bf k}_2,q}_{j,j}\ne {\mathbb 0}$, $\mathcal{P}^{{\rm\bf k}_1,p}_{i,i}\mathcal{P}^{{\rm\bf k}_2,q^\prime}_{j^\prime,j^\prime}\ne {\mathbb 0}$, $\mathcal{P}^{{\rm\bf k}_1,p^\prime}_{i^\prime,i^\prime}\mathcal{P}^{{\rm\bf k}_2,q^\prime}_{j^\prime,j^\prime}\ne {\mathbb 0}$, and so on are also true. The existence of complex link associated with a node at HSPs on the surface of the BZ is a reflection of multiple unitary similarity transformations which block diagonalise the EBR yielding the same direct sum of invariant sub vector space there.

However, the von Neumann-Wigner (anti-crossing) theorem and dimensionality argument would preclude links between more than two nodes/terminals. Complex links need to be resolved into unique link between two terminals at two nodes. These leads to {\em multiple configurations} of connectivity of bands described by an EBR in the atomic limit. The existence of split EBRs\cite{Cano_J:2018A} where its is classified as decomposable\cite{Elcoro_L:2017} is included under these configurations. But other configurations, which are symmetry permitted, may have no such classifications.

The existence of complex links and the need for its resolution means there are multiple ways to block-diagonalise the representation matrix of full EBR at associated HSPs on the surface of BZ. How to implement the block-diagonalisation is not provided in the transformation properties of full vector space spanned by the EBRs. Thus transformation properties of EBR in the literature is incomplete in the description of energy band, and requires further context of band connectivity.

To see the potential of projection operator technique, it can be applied to the configuration of connectivity obtained under compatibility relations shown in Fig.\ref{fig:connected}. For EBR induced from p$_z$ orbitals centred on Wyckoff position 2b of the honeycomb lattice, we have
\begin{align}
\mathcal{P}^{{\rm M},{\rm M}_4^+}_{1,1}\mathcal{P}^{\Gamma,\Gamma_4^+}_{1,1}&=\mathcal{P}^{\Gamma. \Gamma_4^+}_{1,1} \notag \\
\mathcal{P}^{{\rm M},{\rm M}_3^-}_{1,1}\mathcal{P}^{\Gamma,\Gamma_4^+}_{1,1}&=\mathbb{0}. \notag
\end{align}
Thus there is a simple link between $\Gamma_4^+$ and M$_4^+$ with no link between $\Gamma_4^+$ and M$_3^-$. Therefore the only symmetry permitted configuration of connectivity from this EBR is shown in Fig.\ref{fig:connected} (a). Those in Fig.\ref{fig:connected} (b) is symmetry forbidden from this EBR despite its compliance to compatibility relations. Thus it is symmetry indicated as topologically non-trivial.


\subsection{Resolving complex link into unique links and multiple configuration of band connectivity in atomic limit}
\begin{figure}[b]
\includegraphics[width=0.7\textwidth]{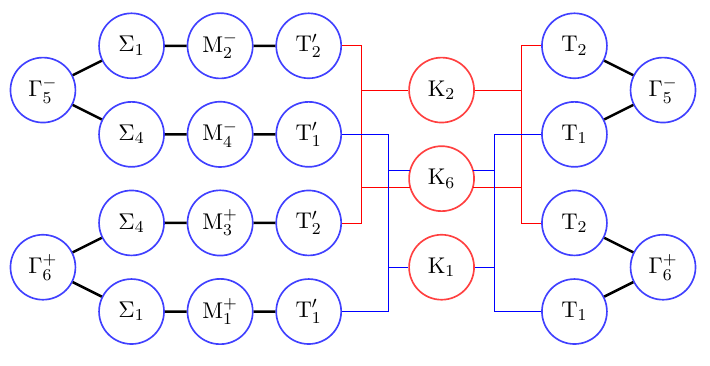}
\caption{Link diagram of EBR from spin-less ${\rm p}_{x,y}$ orbital on Wyckoff position 2b of honeycomb lattice. Simple links are shown in thick lines and complex links are shown in thin lines involving more than 2 nodes/terminals. All links to nodes at K are complex.}
\label{fig:pxy_atomic_limit}
\end{figure}
This is perhaps best illustrated with an example of EBR induced from p$_{xy}$ orbitals centred on Wyckoff position 2b of the honeycomb lattice. The irrep decomposition of the EBR at HSPs are shown in Tab.\ref{tab:spG80_decomp} establishing symmetry labels of the nodes and terminals that exist. Using projection operator technique, the existence of simple and complex links between these nodes are established. All links connected to nodes/terminals at K are complex. All links between $\Gamma$ and M are simple. This is illustrated in Fig.\ref{fig:pxy_atomic_limit}.
\begin{figure}[b]
\includegraphics[width=0.32\textwidth]{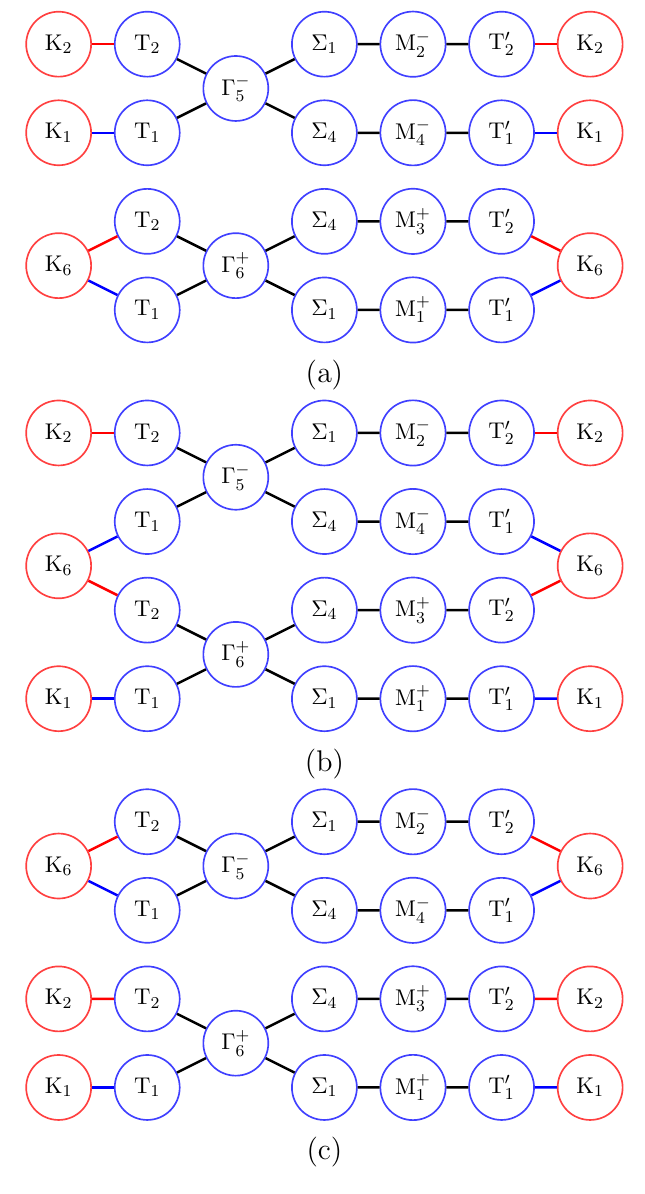}
\includegraphics[width=0.47\textwidth]{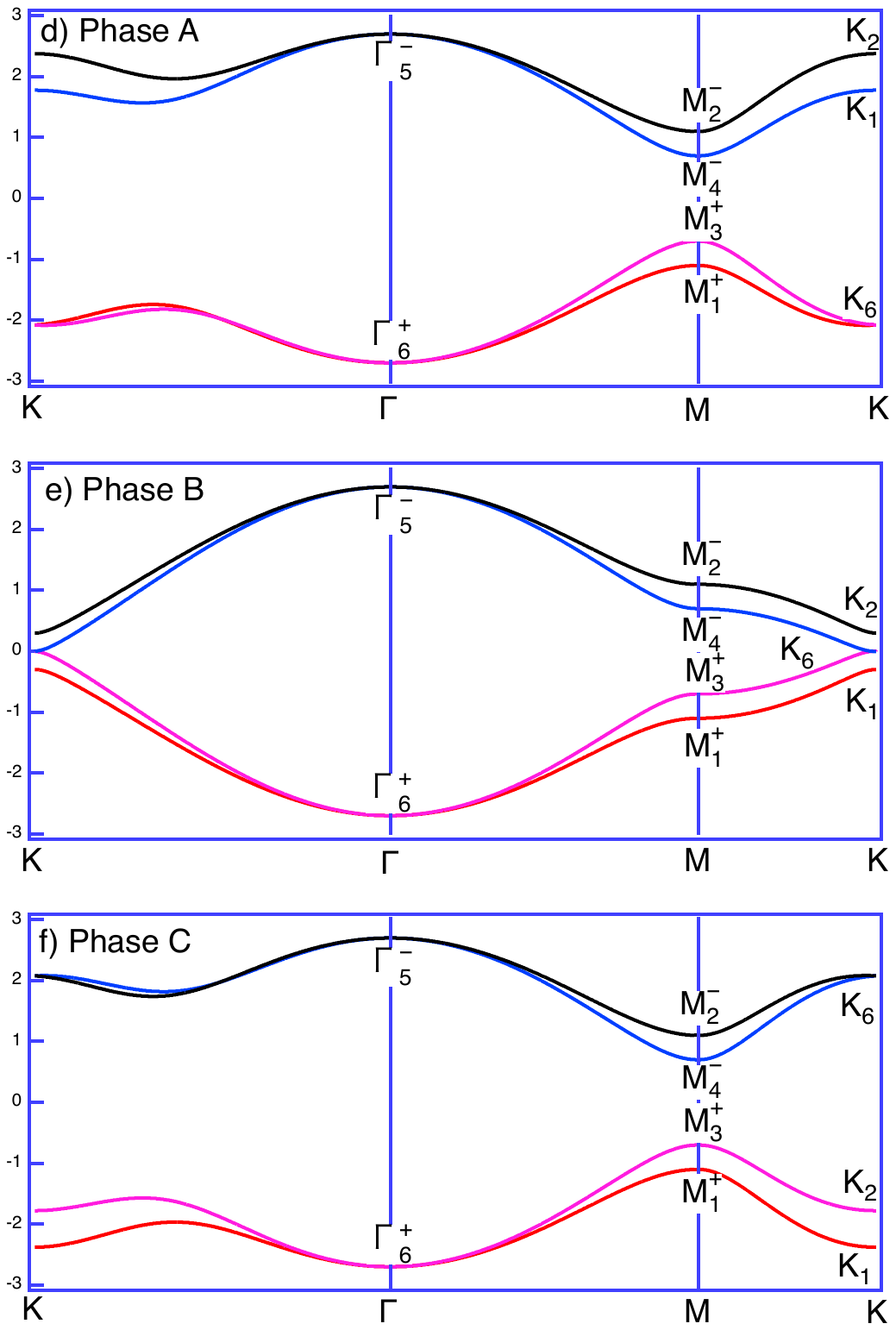}
\caption{\label{fig:pxy}Symmetry compliant band connectivities (a) - (c) obtained from projection operator method, and dispersion relations from EBR of $\Gamma_6$ Wannier functions centred on Wyckoff position 2b ($\mathcal{G}=80, p6/mmm(L))$. The dispersion relations are obtained from TB model using hopping parameters: $t^{FNN}_{\Gamma_6,\Gamma_6:A}=0.9$, $t^{FNN}_{\Gamma_6,\Gamma_6:B}=0.1$, $t^{SNN}_{\Gamma_6,\Gamma_6:A}=0$, $t^{SNN}_{\Gamma_6,\Gamma_6:B}=0$, $t^{SNN}_{\Gamma_6,\Gamma_6:C}=0$ for all there phases; and (d) Phase A: $t^{SNN}_{\Gamma_6,\Gamma_6:D}=0.4$, (e) Phase B: $t^{SNN}_{\Gamma_6,\Gamma_6:D}=0$, and (f) Phase C: $t^{SNN}_{\Gamma_6,\Gamma_6:D}=-0.4$.}
\end{figure}

The von Neumann-Wigner (anti-crossing) theorem and dimensionality argument exclude any links between more than two nodes/terminals. Thus complex link need to be resolved into unique links between two nodes/terminals leading to distinguishable configurations of band connectivity. Some of the configurations are illustrated in Fig.\ref{fig:pxy} together with dispersion relations obtained using symmetry compliant Hamiltonian described in Appendix.\ref{sec:III}. In particular, Fig.\ref{fig:pxy} (a) and (c) are two possible distinguishable configurations with disconnected components and (b) is fully connected. The corresponding dispersion relation from the symmetry compliant TB model are shown in (d), (f), and (e). There are other fully connected configurations not shown in the figure. If one defines all the hopping parameter in terms of a single energy scale, the atomic limit of the connectivity can be obtained by letting the energy scale to approach zero whilst maintaining the relative size of the hopping parameters. Thus three distinct configurations of connectivity can be reached in the atomic limit.

The multiple configurations of band connectivity realised here are due to different ways the vector space spanned in by the EBR at K (where the gauge term is not unity) can be decomposed into direct sums of invariant subspaces. With the {\em exception} of nodes at HSPs on the surface of BZ with non-unity gauge term, the connectivity of any two nodes/terminals of an EBR is fixed (simple link) and determined by Eq.\eqref{eqn:EBRTB_matrix}. At HSPs on the surface of BZ with non-unity gauge term, the choice of different block-diagnoalisation of the vector space spanned by the EBR is made by the relative strength of hopping parameters in a symmetry compliant TB model. In particular, the K$_6$ irrep can arise from three different linear combination of components of the EBRs leading to the three different configurations of connectivity shown in Fig.\ref{fig:pxy}. The results of complex links obtained from projection operator reflect this outcome.

An important consequence is that the configurations of band connectivity in the atomic limit are not unique for EBR induced from Wannier functions centred on the crystallographic orbit of Wyckoff positions with multiplicity. Which of the three different atomic limits shown by the configuration of connectivity in Fig.\ref{fig:pxy} (a)-(c) manifest in band structure clearly depends on {\em relative} hopping parameters before the energy scale is taken to zero, i.e. it is dependent on the dynamics. Associated with the split EBR configurations is the concept of dynamic irreducible BRs which we address in the next section.

It is important to note that fixing connectivity between $\Gamma$ and K does not fix the connectivity between M and K. This implies one can have gap existing in one direction but connected in another. A gapped system requires gap in all directions.

\section{Irreducible BRs and equivalent BRs}
\label{sec:irreducible_br}
The concept of irreducible and equivalent BRs have been discussed in the literature as a properties of the basis\cite{Zak_J:1980, Michel_L:2001A} but rarely in association with the different configuration of band connectivity. Given the existence of multiple configurations of band connectivity of  EBRs in the atomic limit, such concepts clearly need to be qualified by the configuration of connectivity.
\subsection{Definition of irreducibility of BR and equivalence between BRs}
An irreducible band representation (IBR) is defined as a band representation induced from Wannier functions centred on the crystallographic orbits of Wyckoff positions or on a linear combination of a sub-set of such crystallographic orbits on the $\tau_\alpha$ index, which spans invariant vector spaces at all wave-vector ${\rm\bf k}\in$ BZ, for a {\em connected band} under a given configuration of band connectivity. The dimensionality of an IBR must be the same as the connected band described by it, and it describes no other state not among the connected band.

The transformation properties of EBR shown in Eq.\eqref{eqn:EBRTB_matrix} and band connectivity from nodes at $\Gamma$ to any nodes at ${\rm\bf k}$, where the gauge term is unity, are well defined and unique. i.e. all links are simple links under the constraint of unity gauge term. The multiple ways of block diagonalising the EBR at HSPs on the surface of BZ occurs where the gauge term takes non-unity value. This can lead to split configurations of band connectivity and multiple atomic limits. A connected band, if described at every ${\rm\bf k}\in$ BZ by BR induced from Wannier functions centred on the crystallographic orbits of Wyckoff positions or linear combination of such orbits and having the same dimensionality, defines the {\em band irreducibility}. Unlike basis of irrep of point group or SO(3), an IBR basis itself may also describe further split band configuration of band connectivity under a different set of hopping parameter of its own. In this case, one needs to consider new IBRs based on the given configuration of band connectivity. An example is a fully connected EBR (Fig.\ref{fig:pxy}b), which may admit split bands (Fig.\ref{fig:pxy}a\&c) described by IBRs under a different set of hopping parameters. Therefore, the occurrence of IBR depends on the dynamics/topolology of the bands.

Equivalence between two BRs requires transformation properties to be related by a ${\rm\bf k}$ independent similarity transformation for all ${\rm\bf k}\in$ BZ under the same connectivity between the nodes/terminals at HSPs. This definition differs from those in the literature\cite{Bradlyn_B:2017, Cano_J:2018B} which requires continuous evolution from one BR to another based on a parameter. As the connectivity/topology is involved in the definition here (not an intrinsic properties of the basis), it is difficult to reconcile the differences. Here the equivalence is only in terms of transformation properties of bands based on an {\em infinite} solid. It does not imply any other equivalence, such as topological properties. This is made clear in Sec.VII of the accompanying manuscript\cite{Zhang_J:2024}.

Both of these two concepts need to be considered in term of symmetry that manifest itself in real space and its dependence on the configurations of band connectivity. In real space, the band invariant vector spaces are associated with the infinite number of Wannier functions centred on the crystallographic orbits of equivalent Wyckoff positions. Any symmetry operation under the space group $\mathcal{G}$ permute these orbits. One might expect band invariant subspace from linear combination of EBR components. With the knowledge of different atomic limits of EBRs, one can see how irreducible BR may arise from taking the Fourier transform with respect to the positions of these orbits.

First of all, a fully connected EBR is irreducible. A fully connected EBR clearly satisfy the dimensionality requirement and preclude any split band configuration of connectivity. This implies a particular choice is made among the multiple ways in diagonalising the representation matrix at HSPs on the surface of BZ where the gauge term is not unity. This is despite of the permutation symmetry of the crystallographic orbits in real space under the action of $\mathcal{G}$. The expectation of invariant subspace from linear combinations of EBR components is excluded by the choice on full connectivity. For most ${\rm\bf k}$ in the interior of BZ, the vector space spanned by an EBR contains band invariant subspaces arising from linear combination on equivalent Wyckoff position index. Such invariance is broken on HSPs on the surface of BZ where the gauge term in the representation matrix in Eq.\eqref{eqn:EBRTB_matrix} takes non-unity values. The choice of block-diagonalisation of Eq.\eqref{eqn:EBRTB_matrix} made under full connectivity breaks the expected invariance from real space. Thus when the atomic limits are in a fully connected configuration, the EBR is irreducible. This is perhaps the cause of erroneous expectation of fully connected EBRs\cite{Michel_L:2001}

However, the gauge term $\exp\{i{\rm\bf g}_R\cdot{\rm \tau}_{\alpha^\prime}\}$ is also responsible for the existence of complex links associated with these HSPs and the need to resolve such link into unique links in the band connectivity. Such resolution {\em can} lead to disconnected configurations of band connectivity and {\em can restore} the invariance of linear combinations of EBRs at these HSPs under such configuration of band connectivity. The {\em sufficient condition} for restoration of invariance is that the disconnected bands are connected to nodes/terminals at HSP whose irrep labels are obtained with the gauge term $\exp\{i{\rm\bf g}_R\cdot{\rm \tau}_{\alpha^\prime}\}$ absent. In the example shown in Fig.\ref{fig:pxy} (a) and (c), the disconnected component bands are results of such invariant linear combination arising from adding or subtracting on the $\tau_\alpha$ index. Under these particular disconnected configurations, the BR basis of the disconnected components linked to K$_6$ is irreducible. Since the fully connected EBR is band irreducible, the BR basis of the other disconnected components linked to K$_1$ and K$_2$ is also band irreducible. As this band from the configuration of band connectivity is {\em not} equivalent to any EBR centred on Wyckoff position 1a, the basis of symmetry indicator method is re-examined in the accompanying manuscript\cite{Zhang_J:2024}.

These IBR arising from linear combination of EBR components lost the $\tau_\alpha$ index. Thus their transformation properties can be equivalent to some EBR induced from Wannier functions centred on Wyckoff position with no multiplicity. In the example shown in Fig.\ref{fig:pxy}a\&c, the BR basis of connected component linked to node K$_6$ are equivalent to EBRs centred on Wyckoff position 1a. These equivalence of BRs can only be understood in the context of band connectivity. The BRs on either side of the equivalence clearly have different real space representations with associated Wyckoff position 2b and 1a, but their representation matrices are related by a ${\rm\bf k}$ independent similarity transformation. It is difficult to reconcile the equivalence here with those defined in Ref.\onlinecite{Bradlyn_B:2017, Cano_J:2018B}. In contrast, those bands connected to nodes K$_1$ and K$_2$ give rise to IBR with no such equivalence. 

\subsection{Dynamic irreducible BRs among CBRs and dynamic equivalent BRs}
The key in realising irreducible BR from linear combination of EBR components is the restoration of invariance at HSPs on the surface of BZ. When this occurs, the symmetry expected from real space is consistent with the irreducible BRs. However, EBRs induced from s or p$_z$ orbitals centred on Wyckoff position 2b have only a single connected configuration of band connectivity. Indeed, these EBRs are described as indecomposable in the BANDREP\cite{Elcoro_L:2017}. 

How does one reconcile the real space symmetry on the permutation of orbits of Wyckoff position and the lack of split EBRs. The lack of band invariant subspace is clearly frustrated by absence of split configuration of band connectivity in such EBR. Such frustration can be removed through hybrid orbitals/CBRs and interaction. This is best illustrated by the hybridisation between s and p$_{xy}$ orbitals centred on Wyckoff position 2b. They form the very strong $\sigma$ bonds between carbon atoms in graphene. 

\begin{figure}[t]
\includegraphics[width=0.4\textwidth]{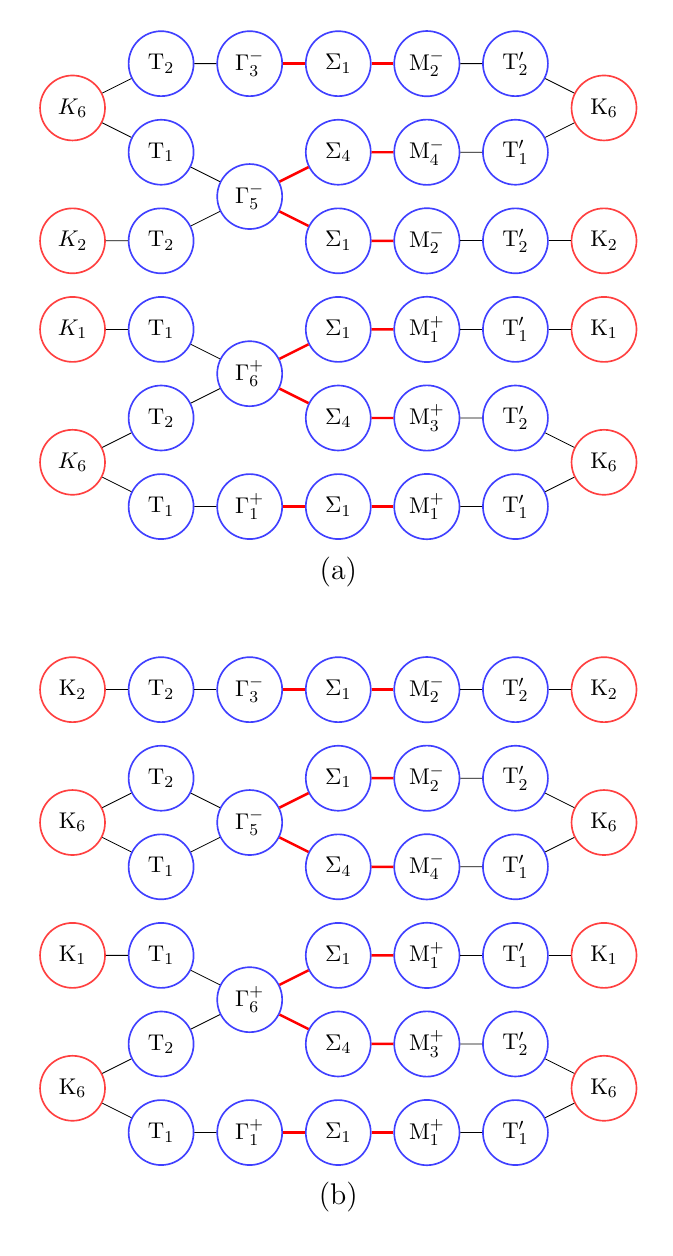} 
\includegraphics[width=0.55\textwidth]{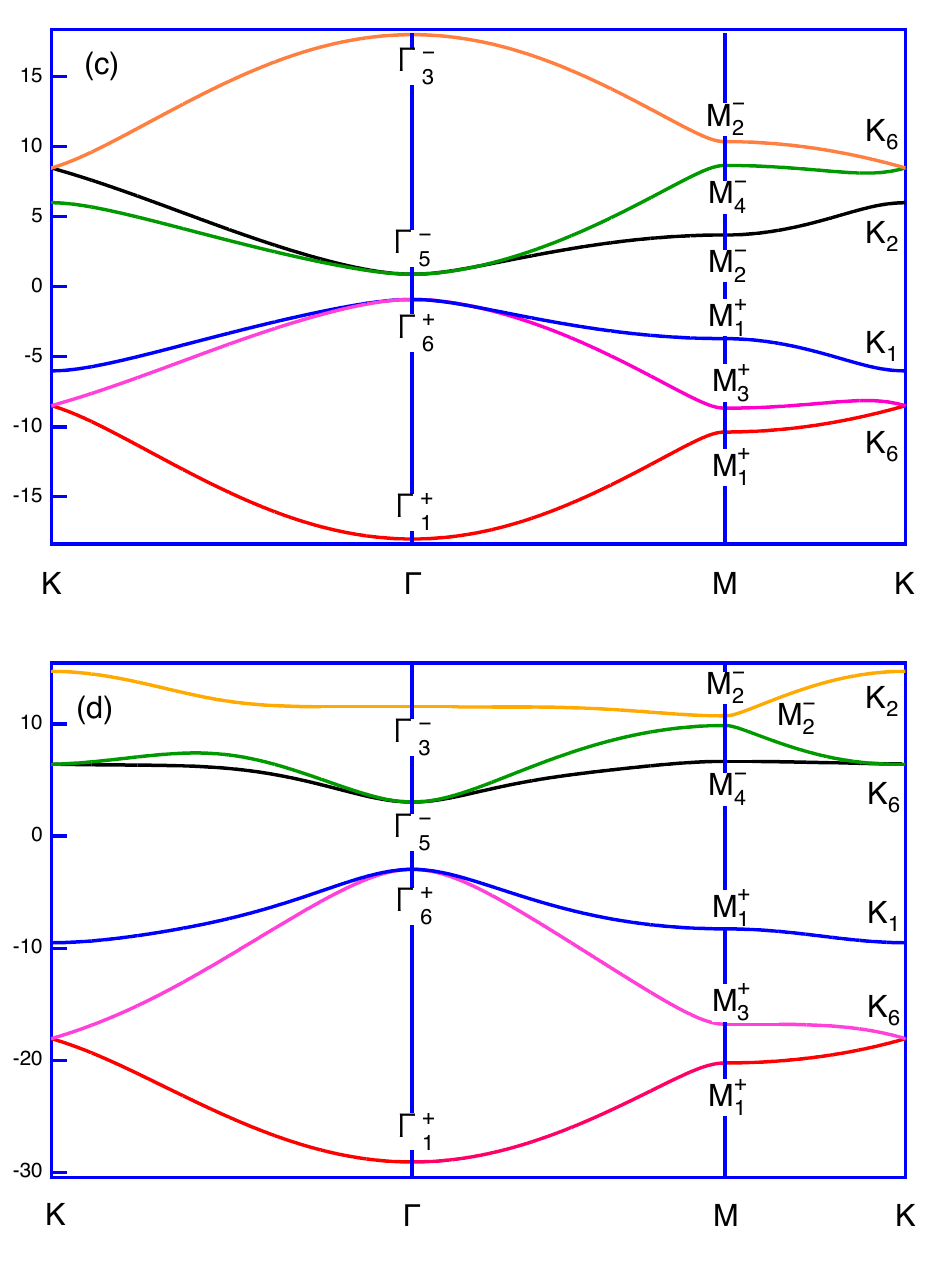}\\
\caption{Symmetry compliant band connectivity of the ${\rm sp_2}$ Wannier functions centred on Wyckoff position 2b. (a) and (b) both exhibit gap at half filling and (b) exhibit additional gap at 5/6 filling. (c) and (d) shows the corresponding dispersion obtained from TB with hooping parameters: (c) $t^{ZNN}_{\Gamma_1,\Gamma_1}=t^{ZNN}_{\Gamma_6,\Gamma_6}=0$, $t^{FNN}_{\Gamma_1,\Gamma_1}=-6$, $t^{FNN}_{\Gamma_1,\Gamma_6}=4$, $t^{FNN}_{\Gamma_6,\Gamma_6:A}=0.3$, $t^{FNN}_{\Gamma_6,\Gamma_6:B}=2$, no second nearest neighbour interaction included. (d) $t^{ZNN}_{\Gamma_1,\Gamma_1}=-8.868$, $t^{ZNN}_{\Gamma_6,\Gamma_6}=0$, $t^{FNN}_{\Gamma_1,\Gamma_1}=-6.769$, $t^{FNN}_{\Gamma_1,\Gamma_6}=5.580$, $t^{FNN}_{\Gamma_6,\Gamma_6:A}=-1.002$, $t^{FNN}_{\Gamma_6,\Gamma_6:B}=4.035$,  $t^{SNN}_{\Gamma_1,\Gamma_1}=0.020$, $t^{SNN}_{\Gamma_1,\Gamma_6:A}=0.050$, $t^{SNN}_{\Gamma_1,\Gamma_6:B}=0.050$, $t^{SNN}_{\Gamma_6,\Gamma_6:A}=0.010$, $t^{SNN}_{\Gamma_6,\Gamma_6:B}=0.210$,  $t^{SNN}_{\Gamma_6,\Gamma_6:C}=0.50i$, $t^{SNN}_{\Gamma_6,\Gamma_6:D}=0.510$. }
\label{fig:sp2}
\end{figure}
Fig.\ref{fig:sp2} shows two possible configurations of band connectivity in (a) and (b) and physical dispersion (c) and (d) obtained from TB model with composite EBRs with s and p$_{xy}$ orbitals centred on orbits of Wyckoff position 2b on the honeycomb lattice. Here the dynamic interaction between the s and p$_{xy}$ orbital allows the restoration of invariance from linear combination of components of induced EBRs at HSPs on the surface of BZ. The original EBR basis can be labelled by the Wyckoff position and the Wannier function irrep  as $\left|{\bm \tau}_A^{2b}, \Gamma_1,1\right>$, $\left|{\bm \tau}_B^{2b}, \Gamma_1,1\right>$, $\left|{\bm \tau}_A^{2b}, \Gamma_6,1\right>$, $\left|{\bm \tau}_A^{2b}, \Gamma_6,2\right>$, $\left|{\bm \tau}_B^{2b}, \Gamma_6,1\right>$,  and $\left|{\bm \tau}_B^{2b}, \Gamma_6,2\right>$. The associated irreducible BRs are identified by the symmetry label at the $\Gamma$ point. They are $\left| \Gamma_1^+,1\right>$, $\left|\Gamma_6^+,1\right>$, $\left| \Gamma_6^+,2\right>$, $\left|\Gamma_3^-,1\right>$, $\left| \Gamma_5^-,1\right>$,  and $\left| \Gamma_5^-,2\right>$. The linear combination can then be established using projection operators in Eq.\eqref{eqn:projection} at $\Gamma$ point from the original basis. They are related to the original EBR basis by the {\em unitary} similarity transform
\begin{equation}
\mathcal{U}=\frac{1}{\sqrt{2}}\left(\begin{array}{rrrrrr}\phantom{\text{-}}1 & 0 & 0 & 1 & 0 & 0 \\ 1 & 0 & 0 & \text{-}1 & 0 & 0 \\ 0 & 1 & 0 & 0 & 1 & 0 \\ 0 & 0 & 1 & 0 & 0 & 1 \\ 0  & 1  & 0 & 0 &\text{-}1 & 0 \\ 0 & 0 & 1 & 0 & 0 & \text{-}1 \end{array}\right).
\label{eqn:sp2}
\end{equation}
Focusing on the connectivity in Fig.\ref{fig:sp2}b, the isolated band connected to $\Gamma_3^-$ is linked to K$_2$ and hence the EBR combination $\left|\Gamma_3^-,1\right>$ is irreducible. The same can be said about $\left| \Gamma_5^-,1\right>$,  $\left| \Gamma_5^-,2\right>$ as they are now connected to K$_6$. Below the gap at half filling, the bands are connected. Assessing the two irreducible BRs $\left| \Gamma_1^+,1\right>\oplus\left|\Gamma_6^+,1\right>\oplus\left| \Gamma_6^+,2\right>$ together, they are now connected to ${\rm K}_1\oplus {\rm K}_6$ and the invariance at K is restored for the direct sum of irreducible basis. 

In Fig.\ref{fig:sp2}a, the connected bands above and below the gap are assessed as associated with $\left|\Gamma_3^-,1\right>\oplus\left| \Gamma_5^-,1\right>\oplus\left| \Gamma_5^-,2\right>$ and $\left| \Gamma_1^+,1\right>\oplus\left|\Gamma_6^+,1\right>\oplus\left| \Gamma_6^+,2\right>$ respectively. Their connections at K point again restore the invariance for the direct sum of irreducible BRs. The appearance of IBRs from linear combination of components of EBR induced from s and p$_z$ orbitals centred on Wyckoff position 2b despite them being labelled indecomposable.  
 
The IBR labelled $\left|\Gamma_3^-,1\right>$ originate from linear combination of components of EBR induced from s orbital centred on Wyckoff position 2b. The latter had been classified as indecomposable in BANDREP. Thus such classification need to be reconsidered under CBR and dynamic interaction. Despite the fact that these dynamic IBRs originate from Wannier functions centred on Wyckoff position with multiplicity, they are all of type 2 IBR and are equivalent to other EBR induced from Wannier functions centred on Wyckoff position without multiplicity. Such equivalence can only occur under the constrained dynamic interaction which preserve the band topology. The IBRs and BR equivalence are crucial concepts in the study of symmetry constraints on Berry phase in the accompanying manuscript\cite{Zhang_J:2024}.

\subsection{Type of IBRs and their transformation properties}
The transformation properties of components of EBR, as given in Eq.\eqref{eqn:EBRTB_matrix}, are independent of band connectivity/topology except at HSPs on the surface of BZ where the gauge term takes non-unity values. With the exception of these HSPs on the surface of BZ, the block-diagonalisaton (decomposition EBR in terms of irreps of $\mathcal{G}$) of representation matrix of EBR are unique and well defined. The basis of eigenstate at $\Gamma$ can be obtained by projection operators and the evolution of such linear combination of eigenstate in ${\rm\bf k}$ space produces the bands. However, the deterministic behaviour is broken at these HSPs on the surface of the BZ associated with non-unity gauge term. The actual transformation properties of a given band or IBR induced from Wannier functions centred on linear combination of crystallographic orbits of Wyckoff positions are then determined by the band connectivity/topology at these HSPs. Based on what occurs at these HSPs, one can classify three types of IBRs. 

The first type is a complete EBR appearing in a set of connected bands. The dimensionality of the vector space spanned by the EBR at each ${\rm\bf k}\in$ BZ equals to the number of bands. The basis of the bands are obtained from projection operator at $\Gamma$ point, and they evolve in ${\rm\bf k}$ to form the band. At HSPs on the surface of BZ with non-unity gauge terms, the transformation property of each of the bands results from allocation of irrep labels at these HSPs based on the configurations of the {\em connected} band connectivity. The whole EBR is an IBR. These can be induced from Wannier functions centred on Wyckoff positions with or without multiplicity. 

The second and third types of IBR arises from Wannier functions centred on the linear combinations of cyrstallographic orbits of Wyckoff position with multiplicity. They both originate from the real space permutation symmetry of crystallographic orbits of equivalent Wyckoff positions with multiplicity. Starting at $\Gamma$ point, they naturally form the eigenstate as they are induced from Wannier functions centred on the invariant sub-space of permutation symmetry of the crystallographic orbits of Wyckoff position with multiplicity. Such IBR evolve in ${\rm\bf k}$ space to produce the bands.  At HSPs on the surface of BZ with non-unity gauge terms, the transformation property of each of the bands results from allocation of irrep labels at these HSPs based on the configuration of the {\em disconnected} band connectivity. For band allocated with irrep labels that is given by dropping the gauge term, then the IBR basis has the same dimensionality as the component band and the band irreducibility is restored. These are classified as type 2. Examples includes those IBR identified from component of split EBR connected to K$_6$ in Fig.\ref{fig:pxy} or isolated band with label $\Gamma_3^-$ in phase (b) of Fig.\ref{fig:sp2} under CBR. Type 2 IBR can be labelled by its irrep label at $\Gamma$ and is always equivalent to some other EBR arising from Wyckoff position with no multiplicity and Wannier function of the same symmetry of the site symmetry group (of Wyckoff position with no multiplicity). Type 2 IBR can be further subdivided into Type 2$\alpha$  and type 2$\beta$. In type 2$\alpha$, the restoration of invariance is achieved within a decomposable EBR with accompanied Type 3 IBR. Type 2$\beta$ occurs when the restoration of invariance can only be achieved through interaction within the CBR context.

Type 3 IBR may arises accompanying type 2$\alpha$ from part of a split decomposable EBR and its direct sum with complementary type 2$\alpha$ IBRs spanned the space of the original EBR. IBR identified in Fig.\ref{fig:pxy} from component bands connected to K$_1$ and K$_2$ are example of type 3 IBRs. The transformation properties of type 3 IBR in the BZ are given by block-diagonalisation of Eq.\eqref{eqn:EBRTB_matrix} except where the gauge term takes non-unity value at HSPs on the surface of BZ. At these HSPs, the transformation properties is determined by the band connectivity which also defined the corresponding type 2$\alpha$ IBR.

\subsection{Split EBRs and band topology}
\label{sec:spinpz}
\begin{figure}[b]
\includegraphics[width=0.8\textwidth]{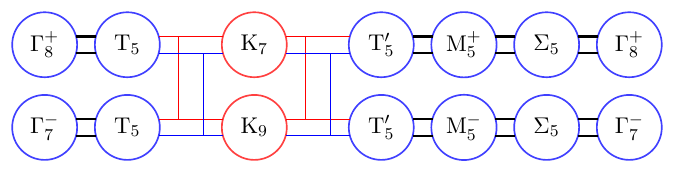}
\caption{Link diagram of EBR induced from spin-full ${\rm p}_z$ orbital on Wyckoff position 2b on honeycomb lattice. Links to nodes at K are complex.}
\label{fig:connected2}
\end{figure}
The last three sub-sections demonstrated possibility of restoring irreducibility of linear combination of EBR components through resolution of complex link or dynamic interaction  in CBRs. However, not all symmetry permitted split EBRs or dynamic interaction among CBR lead to IBRs. Graphene is perhaps the best known example. Instead of restoring irreducibility, the split EBR (PEBR) induced from spin-full p$_z$ orbital centred on Wyckoff position 2b lead to topologically non-trivial phase. 

Using projection operator technique, the link diagram of EBR induced from spin-ful ${\rm p}_z$ orbital ($\mu=\Gamma_8$ of $\tilde{\rm G}^{\bm{\tau}}={\rm D}_{3h}$) is shown in Fig.\ref{fig:connected2}. Any link to nodes/terminals at K are complex and need to be resolved. 
\begin{figure}[t]
\includegraphics[width=0.51\textwidth]{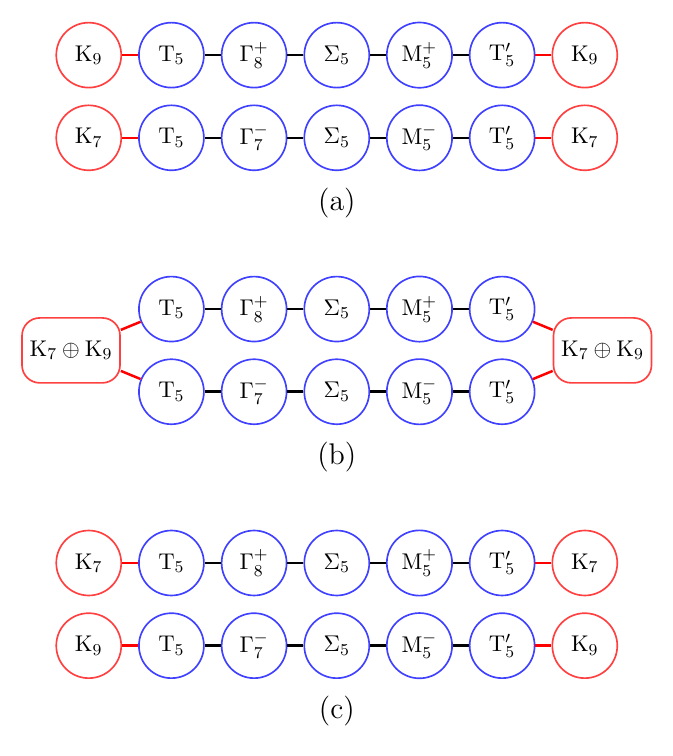}
\includegraphics[width=0.34\textwidth]{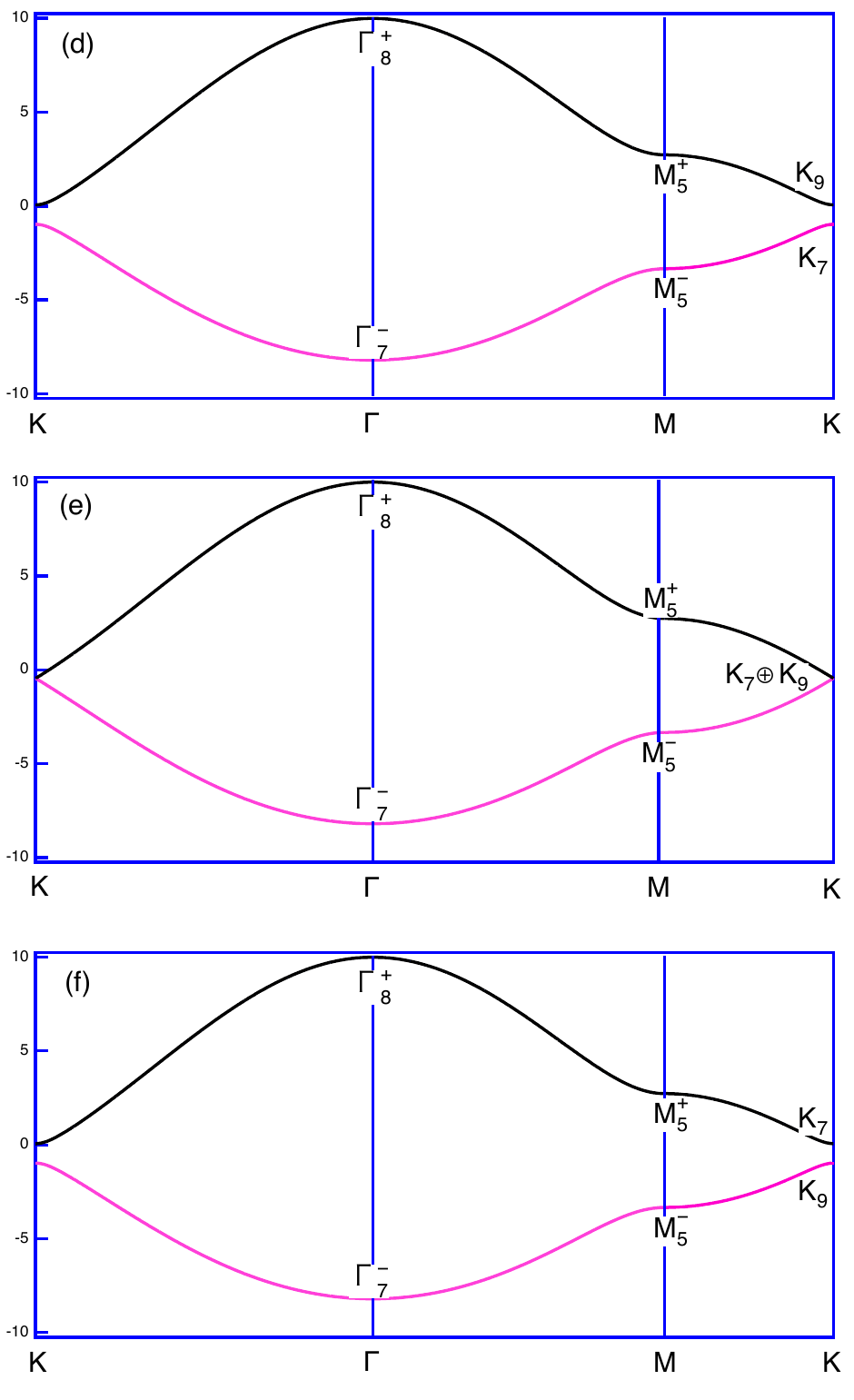}\
\caption{Band connectivities of the spin-full p$_z$ orbital centred on Wyckoff position 2b. (a)-(c) are band connectivities and (d)-(e) are the corresponding dispersion from TB model with the following hopping parameters. $t_{\Gamma_8, \Gamma_8}^{ZNN}=0$, $t_{\Gamma_8, \Gamma_8}^{FNN}=-3.033$, $t_{\Gamma_8, \Gamma_8: A}^{SNN}=0.15$. (d) $t_{\Gamma_8, \Gamma_8:B}^{SNN}=0.1i$, (e) $t_{\Gamma_8, \Gamma_8:B}^{SNN}=0$, and (f) $t_{\Gamma_8, \Gamma_8:B}^{SNN}=-0.1i$.}
\label{fig:pz}
\end{figure}
Fig.\ref{fig:pz} shows three configurations of connectivity with corresponding dispersion obtained from symmetry compliant TB calculation. Time reversal symmetry and inversion symmetry ensure the two terminals/links associated with each nodes coincide with each other. Configuration (a) and (c) correspond to split EBRs whereas (b) correspond to (accidental) degenerate K$_7$ and K$_9$ nodes with connected topology. If the split EBR is to restore the invariance of linear combination of EBR components, one expect at least one of the split band to be connected to K$_8$. This is clearly not the case. It is also easily verified that the eigenstates(nodes) of K$_7$ and K$_9$ contains contribution from both linear combinations of EBR component which give rise to the $\Gamma_8^+$ and $\Gamma_7^-$. This is a manifestation of band inversion commonly associated with topological phase in band theory. Therefore the irreducibility is not restored by the split configuration of connectivity of the EBR. 

It is not possible to describe these disconnected band in terms of an IBRs. In the accompanying manuscript\cite{Zhang_J:2024}, it is shown that the absence of IBR description breaks one of the necessary conditions for selection rules on the generalised Berry phase. $\phi_B$ is not forbidden by the space group symmetry and the band cannot be symmetry guaranteed to be trivial. In other words, it is symmetry indicated to be topologically non-trivial.

\section{Summary}
BR has long being used as basis of TB model. EBR appears to have an elevated status similar to irreps in finite groups, yet the concept of IBR had never been identified clearly or employed in understanding of band topology. The projection operator technique developed in this manuscript establishes a systematic method of obtaining configuration of band connectivity of EBRs permitted by symmetry, particularly when the associated Wyckoff positions have multiplicity. The existence of multiple configurations of band connectivity shows that it is possible to have multiple atomic limits in EBRs induced from Wannier functions centred on crystallographic orbits of such Wyckoff positions with multiplicity. In association with configuration of connectivity, the concepts of IBR and equivalent BRs are clearly defined. The existence of IBRs induced from Wannier functions centred on linear combination orbits of Wyckoff positions has its origin on real space symmetry of the such orbits. They may arise from different atomic limit within an EBR and/or dynamic interaction in the context of CBRs. In particular, their existence is associated with connectivity or topology of the bands. It is not possible to discuss IBRs without the context of band connectivity or topology. In turn, one cannot talk about these concepts without the Hamiltonian. The transformation properties of the three types of IBR enables the analysis of symmetry constraint on Berry phase of connected bands in the accompany manuscript\cite{Zhang_J:2024}.

Where the configuration of band connectivity of EBR established by projection operator technique can lead to split bands, the division are symmetry compliant and may occur along decomposable EBR, or lead to topologically non-trivial phases on either side of the gap. In the former case, the basis of component bands are band irreducible. In the later case disconnected bands cannot be described by IBR. The absence of IBR description corresponds to the concept of band inversion across the gap.

Loosing the equivalent Wyckoff position index, the transformation properties of type 2 IBR from linear combinations are equivalent to EBR induced from Wannier functions centred on Wyckoff position without multiplicity. In addition, type 3 IBRs arising from split EBRs have transformation properties with no such equivalence. Whilst IBRs are the smallest unit of BRs from which a description of band structure can be based under TB model, it is not a necessity for a set of connected band to be described by them. The occurrence of such IBR is dependent on the band topology and dynamic interaction. As will be shown in the accompanying manuscript\cite{Zhang_J:2024}, possessing a set of IBR basis in a set of connected band is a necessary condition for selection rule on its generalised Berry phase. The existence of description based on IBRs is an indication of topologically trivial nature of a set of connected bands. Since the manifestation of IBR is dependent on topology/interaction, symmetry of the EBR basis alone is not able to determine the topology of the band. The existence of IBR with no equivalent EBR implies the premise of symmetry indicator method needs to be reviewed. This naturally leads on to the accompany manuscript\cite{Zhang_J:2024}.

\begin{acknowledgments}
The author wishes to acknowledge the Department of Physics and Institute of Advanced Studies at Tsinghua University for supporting his sabbatical leave during the 2018/19 academic year. Without their support, this work would not have been possible. The author also wishes to thank Professors Bangfen Zhu, Xi Chen,  Zheng Liu, Zhengyu Weng at Tsinghua and Dimitri D Vvedensky, Mr. Syed Hussian at Imperial for guidance and discussions. 
\end{acknowledgments}
\raggedbottom
\pagebreak
\appendix
\section{\label{appI}Some group theoretical concept on the representation of space group}

We need to clarify a number of group theoretical concept on space group ${\mathcal G}$ and its representation. The translation operations form an invariant subgroup ${\mathcal T}$ of the space group. In the left coset decomposition of $\mathcal G$, we have
\begin{equation}
	{\mathcal G}=\{R_1|{\rm\bf v}_1\}\mathcal{T}+ \{R_2|{\rm\bf v}_2\}\mathcal{T}+\cdots+ \{R_i|{\rm\bf v}_i\}\mathcal{T}+\cdots+ \{R_h|{\rm\bf v}_h\}\mathcal{T}.
	\label{eqn:A1}
\end{equation}
Seitz notation is used to denote a typical group element $\{R|{\rm\bf v+t}\}$ where $R$ is the rotational part, ${\rm\bf t}$ is a lattice translation, and ${\rm\bf v}$ is the non-lattice translation associated with $R$. Its action on the position vector yield
\[
\{R|{\rm\bf v+t}\}{\rm\bf r}=R{\rm\bf r}+{\rm\bf v+t}.
\]
Typically one chooses $\{R_1|{\rm\bf v}_1\}=\{E|\bm{0}\}$. The curly typeface is used to indicate infinite group such as space group $\mathcal{G}$ or its invariant subgroup of all translations $\mathcal{T}$. Roman typeface are used to indicate finite groups such as the isogonal point group F, little co-group of ${\rm\bf k}$: $\tilde{\rm G}^{\rm\bf k}$, and site symmetry group: $\tilde{\rm G}^\tau$.

The rotational part of the coset representatives $R_i$ form the isogonal point group ${\rm F}$ which is isomorphic to the quotient group ${\mathcal G}/{\mathcal T}$. The coset representatives generally do not form a group. But in the case of symmorphic space group where all non-lattice translations ${\rm\bf v}_i={\rm\bf 0}$, they form the point group F. 

A set of equivalent Wyckoff positions with multiplicity m have position vector $\bm{\tau}_\alpha$ in the primitive cell with $\alpha=1,\cdots,m$. They form the orbit of the Wyckoff position in the primitive cell. Each position ${\bm \tau}_\alpha$ may be reached from ${\bm\tau}_1$ with some space group operation
\begin{equation}
\{R^{\bm\tau}_\alpha|{\rm\bf v}^{\bm\tau}_\alpha+{\rm\bf t}\}{\bm\tau}_1={\bm\tau}_\alpha \label{eqn:A2}	
\end{equation}
where ${\rm\bf t}$ is a lattice translation determined by $\alpha$. There may be distinct elements which satisfy Eq.\eqref{eqn:A2}, but we make choice of these $\{R^{\bm\tau}_\alpha|{\rm\bf v}^{\bm\tau}_\alpha+{\rm\bf t}\}$ once for each $\bm{\tau}_\alpha$. Each of the position has the ``site symmetry group'' ${\rm G^{{\bm \tau},\alpha}}$ whose members leaves $\bm{\tau}_\alpha$ invariant. ${\rm G^{{\bm \tau},\alpha}}$ forms an invariant subgroup of the space group. These site symmetry groups are conjugate to each other.
\begin{equation}
{\rm G^{{\bm \tau},\alpha}}= \{R^{\bm\tau}_\alpha|{\rm\bf v}^{\bm\tau}_\alpha+{\rm\bf t}\}{\rm G^{{\bm \tau},1}} \{R^{\bm\tau}_\alpha|{\rm\bf v}^{\bm\tau}_\alpha+{\rm\bf t}\}^{-1}	
\end{equation}
Its rotational part (co-group of $\bm \tau$) ${\rm \tilde{G}^{{\bm \tau}, \alpha}}$ also form a subgroup of the isogonal point group ${\rm F}$ of the crystal. ${\rm \tilde{G}^{{\bm \tau}, \alpha}}$ is isomorphic/conjugate to ${\rm \tilde{G}^{\bm \tau}=\tilde{G}^{{\bm\tau}, 1}}$(as per the definition of Wyckoff position). In the left coset decomposition of the isogonal point group ${\rm F}$ with respect to $\tilde{\rm G}^{{\bm \tau}}$, the coset representatives are chosen once and designated as $R^{\bm \tau}_\alpha$. 
\begin{equation}
	{\rm F}=R^{\bm \tau}_1 \tilde{\rm G}^{\bm \tau}+\cdots+ R^{\bm \tau}_\alpha\tilde{\rm G}^{\bm \tau}+\cdots+ R^{\bm \tau}_m\tilde{\rm G}^{\bm \tau},\quad\quad R^{\bm\tau}_1=E\label{eqn:A4}
\end{equation}
where $m$ is the multiplicity of the Wyckoff position. There is a one to one correspondence of $R^{\bm\tau}_\alpha$ in Eq.\eqref{eqn:A2} and Eq.\eqref{eqn:A4}. Hence
\begin{equation}
\tilde{\rm G}^{{\bm \tau},\alpha}= R^{\bm \tau}_\alpha\tilde{\rm G}^{\bm \tau}{R^{\bm \tau}_\alpha}^{-1}	
\end{equation}
Given $R\in {\rm F}$, and $R^{\bm\tau}_\beta$, $RR^{\bm\tau}_\beta\in{\rm F}$. Therefore $RR^{\bm\tau}_\beta$ must be in some unique coset with representative $R^{\bm\tau}_\alpha$ and
\begin{equation}
R^{\bm\tau}_\alpha R^{\bm\tau}_\rightarrow(R, R^{\bm\tau}_\beta)=RR^{\bm\tau}_\beta,\quad\quad R^{\bm{\tau}}_\rightarrow(R, R^{\bm\tau}_\beta)\in\tilde{\rm G}^{\bm\tau}.
\label{eqn:A6}
\end{equation}
i.e. $R^{\bm\tau}_\alpha$ and $R^{\bm\tau}_\rightarrow(R, R^{\bm\tau}_\beta)$ pair are uniquely determined by $R$ and $R^{\bm\tau}_\beta$. Conversely, we can say that given $R$ and $R^{\bm\tau}_\alpha$, $R^{\bm\tau}_\leftarrow(R, R^{\bm\tau}_\alpha)$ and $R^{\bm\tau}_\beta$ are uniquely determined in
\begin{equation}
R^{\bm\tau}_\alpha R^{\bm\tau}_\leftarrow(R, R^{\bm\tau}_\alpha)=RR^{\bm\tau}_\beta,\quad\quad R^{\bm{\tau}}_\leftarrow(R, R^{\bm\tau}_\alpha)\in\tilde{\rm G}^{\bm\tau}\label{eqn:A7}	
\end{equation}

In forming the irreducible space group representation in the conventional way\cite{Bradley_C_J:2010}, the concept of little co-group of ${\rm\bf k}$: was introduced. $\tilde{\rm G}^{\rm\bf k}$ is a subgroup of {\rm F}, whose action preserves the wave vector such that
\begin{equation}
	R{\rm\bf k}\equiv {\rm\bf k}; \mbox{~or~}R{\rm\bf k}={\rm\bf k}+{\rm\bf g}_R;\;\;\;R\in\tilde{\rm G}^{\rm\bf k}
\end{equation}
where ${\rm\bf g}_R$ is a reciprocal lattice vector. The isogonal point group {\rm F} is decomposed as left coset of $\tilde{\rm G}^{\rm\bf k}$
\begin{equation}
	{\rm F}=R_1^{\rm\bf k} \tilde{\rm G}^{\rm\bf k}+\cdots+R_\gamma^{\rm\bf k} \tilde{\rm G}^{\rm\bf k}+\cdots+R_q^{\rm\bf k}\tilde{\rm G}^{\rm\bf k},\label{F_dcomp_gk}
\end{equation}
where there is a one-to-one correspondence between the coset representatives and the arms of star of ${\rm\bf k}$. The coset representatives are chosen once and designated $R^{\rm\bf k}_\gamma$. A space group called the little group of {\rm\bf k}: $\mathcal{G}^{\rm\bf k}$, whose action on Bloch functions leaves the wave vector invariant. It is constructed as
\begin{equation}
\label{eqn:Gk_exp}
	\mathcal{G}^{\rm\bf k}=\{R_{{\rm\bf k},\,1}|{\rm\bf v}_1\}{\mathcal T}+\cdots+ \{R_{{\rm\bf k},\,\beta}|{\rm\bf v}_\beta\}{\mathcal T}+\cdots+ \{R_{{\rm\bf k},\,b}|{\rm\bf v}_b\}{\mathcal T},\quad R_{{\rm\bf k},\beta}\in \tilde{G}^{\rm\bf k}
\end{equation}
$\mathcal{G}^{\rm\bf k}$ is a subgroup of the space group ${\mathcal G}$ which can be expressed as left coset partition of ${\rm G}^{\rm\bf k}$:
\begin{equation}
	{\mathcal G}=\{R^{\rm\bf k}_1|{\rm\bf v}_1\}\mathcal{G}^{\rm\bf k}+\cdots+ \{R ^{\rm\bf k}_\gamma|{\rm\bf v}_\gamma\}\mathcal{G}^{\rm\bf k}+\cdots+ \{R ^{\rm\bf k}_q|{\rm\bf v}_q\}\mathcal{G}^{\rm\bf k}, \label{eqn:SpG_exp}
\end{equation} 
where the same one to one correspondence between the coset representative and the arms in the star of ${\rm\bf k}$ in \eqref{F_dcomp_gk} exist.

\section{Transformation of Elementary Band Representations}
\label{sec:App2}
The crystallographic orbits of an Wyckoff position form an invariant space under the action of elements of space group $\mathcal{G}$. If we place symmetric Wannier functions centred at these orbit and transforming according to irrep of the corresponding site symmetry group, they too form and invariant space under $g\in\mathcal{G}$ and form an infinite dimensional representation of $\mathcal{G}$. Such Wannier function can be expressed as
\begin{equation}
W_{\bm{\tau}_\alpha}^{\mu, i}({\rm\bf u})\equiv W_{\bm{\tau}_\alpha}^{\mu, i}({\rm\bf r}-({\rm\bf t}_\ell+\bm{\tau}_\alpha))
\label{eqn:Wannier_fn}
\end{equation}
where $({\rm\bf t}_\ell+\bm{\tau}_\alpha)$ is the localisation centre, or an entry in the crystallographic orbit of the Wyckoff position $\bm{\tau}_\alpha$. The Wannier function transforms, as function of ${\rm\bf u}={\rm\bf r}-({\rm\bf t}_\ell+\bm{\tau}_\alpha)$, according to component $i$ of irrep $\mu$ of the group $\tilde{\rm G}^{\bm{\tau},\alpha}$. In recognition of the translational periodicity of ${\rm\bf t}_\ell+\bm{\tau}_\alpha$, BRs in ${\rm\bf k}$ space are obtained by taking Fourier transform with respect to the primitive cell location ${\rm\bf t}_\ell$ in Eq.\eqref{eqn:Zak_basis} defining the Zak basis.
It can be easily verified that
\begin{equation}
\Phi^{\rm\bf k}_{\bm{\tau}_\alpha, \mu, i}({\rm\bf r})=\Phi^{\rm\bf k+g}_{\bm{\tau}_\alpha, \mu, i}({\rm\bf r}),
\label{eqn:Zak_gauge}
\end{equation}
where ${\rm\bf g}$ is a reciprocal lattice vector. Current literature often drops the Wyckoff position $\bf{\tau}_\alpha$ explicitly shown in Eq.\eqref{eqn:Wannier_fn} in preference of the use of $\bm{\tau}_\alpha$ in the subscript. 

An alternative convention takes the Fourier transform with respect to the localisation centre variable ${\rm\bf t}_\ell+{\bm \tau}_\alpha$ leading to the definition of TB basis in Eq.\eqref{eqn:TB_basis}. This is the normal basis used in the Slater-Koster formulation\cite{Slater_J:1954} of tight binding theory. It is referred to as TB basis in this manuscript. It is clear that the two basis are related by a ${\rm\bf k}$ dependent gauge transformation shown in Eq.\eqref{eqn:gauge_transform}.
The TB basis does not obey Eq.\eqref{eqn:Zak_gauge} and such deficiency can be overcome in the transformation matrices.

The transformation properties of EBR (Zak basis) were derived in the literature\cite{Evarestov_RA:2007, Cano_J:2018B} but their presentation is not helpful with analysis based on its subduced representation of $\mathcal{G}^{\rm\bf k}$ because of explicit ${\rm\bf k}$ dependence on the rotational part. Here an alternative derivation is presented which leads to immediate separation of translational part of a space group element and those related to rotational part for the TB basis. This facilitate that decomposition of EBR at HSPs and development of projection operators. 

To derive the correct transformation property of Eq.\eqref{eqn:Wannier_fn} and Eq.\eqref{eqn:Zak_basis}, Eq.\eqref{eqn:TB_basis}, we invoke the numerical relation that exist for action of space group element $g\in\mathcal{G}$ on a scalar function $g\bullet\psi({\rm\bf r})=\psi(g^{-1}{\rm\bf r})$. Consider a general space group element $\{R|{\rm\bf v+t}\}$. Hence $\{R|{\rm\bf v+t}\}^{-1}=\{R^{-1}|-R^{-1}({\rm\bf v+t})\}$. The the action of such group element on the Wannier function in Eq.\eqref{eqn:Wannier_fn} centred at ${\rm\bf t}_\ell+\bm{\tau}_\alpha$ yield
\begin{align*}
\hat{S}(\{R|{\rm\bf v+t}\})W_{\bm{\tau}_\alpha}^{\mu, i}({\rm\bf r}-({\rm\bf t}_\ell+\bm{\tau}_\alpha))&=W_{\bm{\tau}_\alpha}^{\mu, i}(\{R^{-1}|-R^{-1}({\rm\bf v+t})\}{\rm\bf r}-({\rm\bf t}_\ell+{\bm{\tau}_\alpha}))\\
&=W_{\bm{\tau}_\alpha}^{\mu, i}(R^{-1}[{\rm\bf r}-\{R|{\rm\bf v+t}\}({\rm\bf t_\ell}+{\bm{\tau}_\alpha})]).
\end{align*}
Since $\{R|{\rm\bf v+t}\}$ is a general space group element, 
\begin{equation}
\{R|{\rm\bf v+t}\}({\rm\bf t_\ell}+{\bm{\tau}_\alpha})={\rm\bf t}_{\ell^\prime}+{\bm{\tau}_\beta},
\label{eqn:munumap}
\end{equation}
where $\bm{\tau}_\beta$ is among the set of equivalent Wyckoff positions in the primitive cell. Then
\begin{align}
{\rm\bf t}_\ell&=\{R^{-1}|-R^{-1}({\rm\bf v+t})\}({\rm\bf t}_{\ell^\prime}+{\bm{\tau}_\beta})-{\bm{\tau}_\alpha}\notag \\
&=R^{-1}({\rm\bf t}_{\ell^\prime}+{\bm{\tau}_\beta})-R^{-1}({\rm\bf v+t})-{\bm{\tau}_\alpha}\label{eqn:tellmap}
\end{align}
Hence 
\[
\hat{S}(\{R|{\rm\bf v+t}\})W_{\bm{\tau}_\alpha}^{\mu, i}({\rm\bf r}-({\rm\bf t}_\ell+\bm{\tau}_\alpha))=W_{\bm{\tau}_\alpha}^{\mu, i}(R^{-1}[{\rm\bf r}-({\rm\bf t}_{\ell^\prime}+{\rm\bf \tau_\beta})]).
\]
Treating $W^{\mu, i}_{\bm{\tau}_\alpha}(R^{-1}[{\rm\bf r}-({\rm\bf t}_{\ell^\prime}+{\bm{\tau}_\beta})])=\hat{U}(R)W^{\mu, i}_{\bm{\tau}_\alpha}({\rm\bf u})$ where ${\rm\bf u}={\rm\bf r}-({\rm\bf t}_{\ell^\prime}+{\bm{\tau}_\beta})$, the linear operator $\hat{U}(R)$ can be decomposed into product of three operators according to Eq.\eqref{eqn:A7}
\[
\hat{U}(R)=\hat{U}(R_\beta)\hat{U}(R^\prime)\hat{U}(R_\alpha^{-1}),\quad R^\prime\in\tilde{\rm G}^\tau.
\]
Define
\[
\hat{U}(R_\alpha)W_{\bm{\tau}_1}^{\mu,i}({\rm\bf u})\equiv W_{\bm{\tau}_\alpha}^{\mu,i}({\rm\bf u}),\quad\mbox{hence}\quad\hat{U}(R_\alpha^{-1})W_{\bm{\tau}_\alpha}^{\mu,i}({\rm\bf u})=W_{\bm{\tau}_1}^{\mu,i}({\rm\bf u})
\]
we obtain
\begin{align*}
\hat{U}(R)W^{\mu, i}_{\bm{\tau}_\alpha}({\rm\bf u})&=\hat{U}(R_\beta)\hat{U}(R^\prime)\hat{U}(R_\alpha^{-1})W^{\mu, i}_{\bm{\tau}_\alpha}({\rm\bf u})\\
&=\hat{U}(R_\beta)\hat{U}(R^\prime)W_{\bm{\tau}_1}^{\mu,i}({\rm\bf u})=\hat{U}(R_\beta)\sum_jD^\mu(R^\prime)_{ji}W_{\bm{\tau}_1}^{\mu,j}({\rm\bf u})\\&=\sum_jD^\mu(R^\prime)_{ji}W_{\bm{\tau}_\beta}^{\mu,j}({\rm\bf u})=\sum_jD^\mu(R^\prime)_{ji}W_{\bm{\tau}_\beta}^{\mu,j}({\rm\bf r}-({\rm\bf t}_{\ell^\prime}+\bm{\tau}_\beta))
\end{align*}
where $D^\mu(R^\prime)$ is the representation matrix describing the transformation of $W_{\bm{\tau}_1}^{\mu,i}({\rm\bf u})$ under the action of $\hat{U}(R^\prime), R^\prime\in\tilde{\rm G}^{\tau}$. Substituting this into action of $\hat{S}(\{R|{\rm\bf v+t}\})$, we obtain
\begin{subequations}
\label{eqn:wannier_tfm}
\begin{equation}
\hat{S}(\{R|{\rm\bf v+t}\})W^{\mu, i}_{\bm{\tau}_\alpha}({\rm\bf r}-({\rm\bf t}_\ell+\bm{\tau}_\alpha))=\sum_jD^\mu(R^\prime)_{j,i}W_{\bm{\tau}_\beta}^{\mu,j}({\rm\bf r}-({\rm\bf t}_{\ell^\prime}+\bm{\tau}_\beta)),\quad R^\prime=R_\beta^{-1}RR_\alpha\in\tilde{\rm G}^\tau \label{eqn:wannier_tfma}
\end{equation}
We can define the following representation matrices
\begin{equation}
D^{\bm{\tau}}(R)_{\bm{\tau}_\alpha^\prime, \bm{\tau}_\alpha} =\left\{\begin{array}{ll}1, & \mbox{if~~}{\{R|{\rm\bf v}\}\bm{\tau}_\alpha=\bm{\tau}_\alpha^\prime+{\rm\bf t}^\prime}\\ 0 &\mbox{otherwise}\end{array} \right.,\quad\quad\quad D^{\bm{\tau},\mu}(R)=D^{\bm{\tau}}(R)\overline{\otimes} D^{\mu}(R^\prime).  \label{eqn:atomic_rep}
\end{equation}
Here $\overline{\otimes}$ {\em is an abuse of notation} as $R^\prime$ is dependent on (function of) the row/colunb index of $D^{\bm{\tau}}(R)$. Then Eq.\eqref{eqn:wannier_tfma} can be written in terms of the representation matrix:
\begin{equation}
\hat{S}(\{R|{\rm\bf v+t}\})W^{\mu, i}_{\bm{\tau}_\alpha}({\rm\bf r}-({\rm\bf t}_\ell+\bm{\tau}_\alpha))=\sum_{\bm{\tau}_\beta,j}D^{\bm{\tau},\mu}(R)_{\bm{\tau}_\beta,j; \bm{\tau}_\alpha,i}W_{\bm{\tau}_\beta}^{\mu,j}({\rm\bf r}-({\rm\bf t}_{\ell^\prime}+\bm{\tau}_\beta)).\label{eqn:wannier_tfmb}
\end{equation}
\end{subequations}
Recognising the relation in Eq.\eqref{eqn:tellmap}, we can express 
\[
\exp\{i{\rm\bf k}\cdot{\rm\bf t}_\ell\}=\exp\{iR{\rm\bf k}\cdot{\rm\bf t}_{\ell^\prime}\}\exp\{-iR{\rm\bf k}\cdot({\rm\bf v+t})\}\exp\{-i{\rm\bf k}\cdot\bm{\tau}_\alpha\}\exp\{iR{\rm\bf k}\cdot\bm{\tau}_\beta\}
\]
Utilising the transformation of Wannier functions in Eq.\eqref{eqn:wannier_tfm}, we obtain the transformation properties of the BR defined in Eq.\eqref{eqn:Zak_basis} as:

\begin{subequations}
\label{eqn:tfm}
\begin{align}
&\hat{S}(\{R|{\rm\bf v+t}\})\Phi^{\rm\bf k}_{\bm{\tau}_\alpha, \mu, i}({\rm\bf r})=\exp\{-iR{\rm\bf k}\cdot({\rm\bf v+t})\}\exp\{i(R{\rm\bf k})\cdot\bm{\tau}_\beta\}\exp\{-i{\rm\bf k}\cdot\bm{\tau}_\alpha\} \notag \\
&\quad\quad\quad\quad\quad\quad\quad\quad\quad\quad\quad\times\sum_jD^\mu(R^\prime)_{j,i}\Phi^{R{\rm\bf k}}_{\bm{\tau}_\beta, \mu, j}({\rm\bf r}),\quad R^\prime=R_\beta^{-1}RR_\alpha\in\tilde{\rm G}^\tau\label{eqn:tfma}\\
&=\exp\{-iR{\rm\bf k}\cdot({\rm\bf v+t})\}\sum_{\bm{\tau}_\beta,j}\exp\{i(R{\rm\bf k})\cdot\bm{\tau}_\beta\}\exp\{-i{\rm\bf k}\cdot\bm{\tau}_\alpha\}D^{\bm{\tau},\mu}(R)_{\bm{\tau}_\beta, j;\bm{\tau}_\alpha,i} \Phi^{R{\rm\bf k}}_{\bm{\tau}_\beta, \mu, j}({\rm\bf r})\label{eqn:tfmb}
\end{align}
\end{subequations}
The equivalent for TB basis is given by
\begin{subequations}
\label{eqn:tfmTB}
\begin{align}
\hat{S}(\{R|{\rm\bf v+t}\})\phi^{\rm\bf k}_{\bm{\tau}_\alpha, \mu, i}({\rm\bf r})&=\exp\{-iR{\rm\bf k}\cdot({\rm\bf v+t})\}\sum_jD^\mu(R^\prime)_{j,i}\phi^{R{\rm\bf k}}_{\bm{\tau}_\beta, \mu, j}({\rm\bf r}),\label{eqn:tfmTBa}\\
&=\exp\{-iR{\rm\bf k}\cdot({\rm\bf v+t})\}\sum_{\bm{\tau}_\beta,j}D^{\bm{\tau},\mu}(R)_{\bm{\tau}_\beta,j; \bm{\tau}_\alpha,i}\phi^{R{\rm\bf k}}_{\bm{\tau}_\beta, \mu, j}({\rm\bf r}).\label{eqn:tfmTBb}
\end{align}
\end{subequations}
Note that $R{\rm\bf k}$ may not be one of the arms of $\{\ast{\rm\bf k}\}$. When this happens, $R{\rm\bf k}={\rm\bf k}^\prime-{\rm\bf g}_R$ where ${\rm\bf k},{\rm\bf k}^\prime\in\{\ast{\rm\bf k}\}$. For the Zak basis, we can replace $\Phi^{R{\rm\bf k}}_{\bm{\tau}_\beta, \mu, j}({\rm\bf r})$ with $\Phi^{{\rm\bf k}}_{\bm{\tau}_\beta, \mu, j}({\rm\bf r})$ because of Eq.\eqref{eqn:Zak_gauge}. For the TB basis, additional term in the representation matrix is required to ensure periodicity in ${\rm\bf k}$ space. The representation matrix for the Zak basis are given by
\begin{subequations}
\begin{align}
&\Gamma_{\tilde{\rm G}^{\bm{\tau}}\uparrow}^{\mathcal{G}}(\{R|{\rm\bf v+t}\})_{{\rm\bf k}^\prime, \bm{\tau}_\alpha^\prime, \mu^\prime, i^\prime; {\rm\bf k}, \bm{\tau}_\alpha, \mu, i}=\left<{\rm\bf k}^\prime, \bm{\tau}_\alpha^\prime, \mu^\prime, i^\prime \mid \hat{S}(\{R|{\rm\bf v+t}\}\mid {\rm\bf k}, \bm{\tau}_\alpha, \mu, i\right>\notag \\
&\quad\quad=\exp\{-iR{\rm\bf k}\cdot({\rm\bf v+t})\} D^{\rm\bf k}(R)_{{\rm\bf k}^\prime, {\rm\bf k}}\underbrace{\exp\{i{\rm\bf g}_R\cdot\bm{\tau}_\alpha^\prime\}\exp\{i{\rm\bf k}^\prime\cdot\bm{\tau}_\alpha^\prime\}\exp\{-i{\rm\bf k}\cdot\bm{\tau}_\alpha\} \delta_{\mu^\prime,\mu}D^{\bm{\tau},\mu}(R)}_{\Gamma_\phi}.
\label{eqn:EBR_matrix}
\end{align}
where ${\rm\bf k},{\rm\bf k}^\prime\in\{\ast{\rm\bf k}\}$ and
\begin{equation}
	D^{\rm\bf k}(R)_{{\rm\bf k}^\prime, {\rm\bf k}}=\left\{ \begin{array}{ll}1,& \mbox{if~~}{\rm\bf k}^\prime= R{\rm\bf k}-{\rm\bf g}_R \\ 0 & \mbox{otherwise}\end{array}\right.
	\label{eqn:g_R}
\end{equation}
\end{subequations}
${\rm\bf g}_R$ is an $R$ dependent reciprocal lattice vector and can only take non-zero value on HSPs on the surface of BZ. Similar representation matrix for the TB basis are:
\begin{align}
&\Gamma_{\tilde{\rm G}^{\bm{\tau}}\uparrow}^{\mathcal{G}}(\{R|{\rm\bf v+t}\})_{{\rm\bf k}^\prime, \bm{\tau}_\alpha^\prime, \mu^\prime, i^\prime; {\rm\bf k}, \bm{\tau}_\alpha, \mu, i}=\left<{\rm\bf k}^\prime, \bm{\tau}_\alpha^\prime, \mu^\prime, i^\prime \mid \hat{S}(\{R|{\rm\bf v+t}\}\mid {\rm\bf k}, \bm{\tau}_\alpha, \mu, i\right>\notag \\
&\quad\quad=\exp\{-iR{\rm\bf k}\cdot({\rm\bf v+t})\} D^{\rm\bf k}(R)_{{\rm\bf k}^\prime, {\rm\bf k}}\underbrace{\exp\{i{\rm\bf g}_R\cdot\bm{\tau}_\alpha^\prime\} \delta_{\mu^\prime,\mu}D^{\bm{\tau},\mu}(R)_{{\bm \tau}_\alpha^\prime i^\prime, {\bm \tau}_\alpha i}}_{\Gamma_\phi}.
\label{eqn:EBRTB_matrix}
\end{align}
These matrices for the TB basis are particular simple, with no explicit ${\rm\bf k}$ dependence on the terms labelled $\Gamma_\phi$, This is particularly useful for discussion on the transformation properties of path integrals in the BZ in the accompanying manuscript.

Under this derivation, no assumption is made on how the vector space spanned at a particular ${\rm\bf k}$ may be decomposed into invariant sub-spaces. In particular, there may be multiple ways to decompose the vector space spanned by the EBR yielding the same direct sums of invariant vector spaces at HSPs on the surface of of BZ where the {\em gauge term} $\exp\{i{\rm\bf g}_R\cdot\bm{\tau}_\alpha^\prime\}$ takes non-unity value. Therefore, Eq.\eqref{eqn:EBRTB_matrix} is incomplete in the description of bands, and the transformation properties at these HSPs require further information of configuration of band connectivity. When the band connectivity is split, the EBR can be split along IBRs on either side of the gap. The corresponding EBRs are separate into invariant subspaces of IBRs.

\section{Decomposition of EBRs and IBRs at HSPs on a honeycomb lattice}
\label{sec:App3}
\begin{table} [t]
	\caption{\label{tab:G80-op}Correspondence of elements/conjugacy classes of little co-group of ${\rm\bf k}: \tilde{\rm G}^{\rm\bf k}$ at HSPs, site symmetry group: $\tilde{\rm G}^{\tau}$ of Wyckoff position 2b, 3d of honeycomb lattice with sub-periodic layer group $\mathcal{G}=80({\rm L})$. The operation of the point group F and $\tilde{\rm G}^{1a}$ (not listed) are the same as $\tilde{\rm G}^{\Gamma}$.}
	\begin{tabular}{|c|c||c|c|c|c|c|c|c|c|c|c|c|c|c|c|c|c|c|c|} \hline
	\multirow{6}{*}{$\tilde{\rm G}^{\rm\bf k}$} & $\Gamma ({\rm D_{6h}})$ & $\{E\}$ & $\{2C_6\}$ & $\{2C_3\}$ & $\{C_2\}$ & $\{3C_2^\prime\}$ & $\{3C_2^{\prime\prime}\}$ & $\{\imath\}$ & $\{2S_3\}$ & $\{2S_6\}$ & $\{\sigma_h\}$ & $\{3\sigma_d\}$ & $\{3\sigma_v\}$ \\ \cline{2-14}
	& ${\rm T (C_{2v})}$ & $\{E\}$ & - & - & - & - & $\{C_2\}$ & - & - & - & $\{\sigma_v\}$ & $\{\sigma_v^\prime\}$ & - \\ \cline{2-14}
	& ${\rm K (D_{3h})}$ & $\{E\}$ & - & \cellcolor{blue!25}$\{2C_3\}$ & - & - & \cellcolor{blue!25}$\{3C_2\}$ & - & \cellcolor{blue!25}$\{2S_3\}$ & - & $\{\sigma_h\}$ & \cellcolor{blue!25}$\{3\sigma_v\}$ & - \\ \cline{2-14}
	& ${\rm T^\prime (C_{2v})}$ & $\{E\}$ & - & - & - & - & \cellcolor{blue!25}$\{C_2\}$ & - & - & - & $\{\sigma_v\}$ & \cellcolor{blue!25}$\{\sigma_v^\prime\}$ & - \\ \cline{2-14}
	& ${\rm M (D_{2h})}$ & $\{E\}$ & - & - & \cellcolor{blue!25}$\{C_2\}$ & $\{C_2^\prime\}$ & \cellcolor{blue!25}$\{C_2^{\prime\prime}\}$ & \cellcolor{blue!25}$\{\imath\}$ & - & - & $\{\sigma_h\}$ & \cellcolor{blue!25}$\{\sigma_v\}$ & $\{\sigma_v^\prime\}$ \\ \cline{2-14}
	& $\Sigma {\rm (C_{2v})}$ &  $\{E\}$ & - & - & - & $\{C_2\}$ & - & - & - & - & $\{\sigma_v\}$ & - & $\{\sigma_v^\prime\}$ \\ \hline
		$\tilde{\rm G}^{2b}$ & ${\rm \tilde{G}^{\bf 2b} (D_{3h})}$ & $\{E\}$ & - & $\{2C_3\}$ & - & $\{3C_2\}$ & - & - & $\{2S_3\}$ & - & $\{\sigma_h\}$ & - & $\{3\sigma_v\}$  \\ \hline
		$\tilde{\rm G}^{3d}$ & ${\rm \tilde{G}^{\bf 3d} (D_{2h})}$ & $\{E\}$ & - & - &$\{C_2\}$ & $\{C_2^\prime\}$ & $\{C_2^{\prime\prime}\}$ & $\{\imath\}$ & - & - & $\{\sigma_v\}$ & $\{\sigma_v^\prime\}$ & $\{\sigma_v^{\prime\prime}\}$  \\ \hline

	\end{tabular}
	\label{tbl:grp_map}
\end{table}
Here we perform a decomposition analysis of band representations induced from various Wannier functions centred on Wyckoff position 2b (including ${\rm p}_z$), 3d and 1a on a honeycomb lattice with symmorphic sub-periodic layer group $\mathcal{G}=80$(L)\cite{ITFCE}. The choice of layer group 80 naturally prohibit the Rashba term which breaks inversion symmetry. The Rashba term is not an integral part of the honeycomb lattice and obscure some of the symmetry analysis possible under layer group 80. 
\begin{table} [t]
	\caption{Characters $\chi_\phi$ in Eq.\eqref{eqn:band_decomp} of the EBRs on restriction to $\mathcal{G}^{\rm\bf k}$. The Wannier functions are centred on Wyckoff positions 2b, 3d  of honeycomb lattice with $\mathcal{G}=80({\rm L})$. Where the gauge term $\exp\{i{\rm\bf g}_R\cdot\tau_{\alpha^\prime}\}$ modifies $\chi_\phi$ on HSPs on the surface of BZ, the modified values are given in bracket. For EBR associated with Wyckoff position 2b, only the K point in RD is affected by the gauge term. For EBR associated with Wyckoff position 3d, only the M point in RD is affected by the gauge term.\label{tab:G80}}
	\begin{tabular}{|c|c||c|c|c|c|c|c|c|c|c|c|c|c|c|c|c|c|c|c|} \hline
	\multicolumn{2}{|c||}{EBR} & \multirow{2}{*}{\rotatebox[origin=c]{90}{$\{E\}$}} & \multirow{2}{*}{\rotatebox[origin=c]{90}{$\{2C_6\}$}} & \multirow{2}{*}{\rotatebox[origin=c]{90}{$\{2C_3\}$}} & \multirow{2}{*}{\rotatebox[origin=c]{90}{$\{C_2\}$}} & \multirow{2}{*}{\rotatebox[origin=c]{90}{$\{3C_2^\prime\}$}} & \multirow{2}{*}{\rotatebox[origin=c]{90}{$\{3C_2^{\prime\prime}\}$}} & \multirow{2}{*}{\rotatebox[origin=c]{90}{$\{\imath\}$}} & \multirow{2}{*}{\rotatebox[origin=c]{90}{$\{2S_3\}$}} & \multirow{2}{*}{\rotatebox[origin=c]{90}{$\{2S_6\}$}} & \multirow{2}{*}{\rotatebox[origin=c]{90}{$\{\sigma_h\}$}} & \multirow{2}{*}{\rotatebox[origin=c]{90}{$\{3\sigma_d\}$}} & \multirow{2}{*}{\rotatebox[origin=c]{90}{$\{3\sigma_v\}$}} \\ \cline{1- 2}
	${\bm \tau}$ & $\mu$ of $\tilde{\rm G}^{\tau}$ & & & & & & & & & & & & \\ \hline
		\multirow{7}{*}{2b} & \cellcolor{yellow!25}$\Gamma_4$(p$_z$) & 2 & - & 2 (-1) & - & -2 & - & - & -2 (1) & - & -2 & - & 2 \\ \cline{2-14}
		& \cellcolor{yellow!25}$\Gamma_5$ & 4 & - & -2 (1) & - & 0 & - & - & 2 (-1) & - & -4 & - & 0 \\ \cline{2-14}
		& \cellcolor{green!25}$\Gamma_1$(s) & 2 & - & 2 (-1) & - & 2 & - & - & 2 (-1) & - & 2 & - & 2 \\ \cline{2-14}
		& \cellcolor{green!25}$\Gamma_6$(p$_{x,y}$) & 4 & - &  -2 (1) & - & 0 & - & - & -2 (1) & - & 4 & - & 0 \\ \cline{2-14}
		& $\Gamma_7$ & 4 & - & 2 (-1) & - & 0 & - & - & $2\sqrt{3}$(-$\sqrt{3}$) & - & 0 & - & 0 \\ \cline{2-14}
		& $\Gamma_8$(p$_z$) & 4 & - & 2 (-1) & - & 0 & - & - & -$2\sqrt{3}$($\sqrt{3}$) & - & 0 & - & 0 \\ \cline{2-14}
		& $\Gamma_9$ & 4 & - & -4 (2) & - & 0 & - & - & 0 & - & 0 & - & 0 \\ \hline \cline{2-14}
		\multirow{6}{*}{3d} & \cellcolor{green!25} $\Gamma_1^+$ & 3 & - & - & 3(-1) & 1 & 1 & 3(-1) & - & - & 3 & 1 & 1 \\ \cline{2-14}
		& \cellcolor{green!25} $\Gamma_2^-$ & 3 & - & - & -3(1) & 1 & -1 & -3(1) & - & - & 3 & -1 & 1 \\ \cline{2-14}
		& \cellcolor{yellow!25}$\Gamma_4^+$ & 3 & - & - & -3(1) & -1 & 1 & 3(-1) & - & - & -3 & -1 & 1 \\ \cline{2-14}
		& \cellcolor{yellow!25} $\Gamma_3^-$ & 3 & - & - & 3(-1) & -1 & -1 & -3(1) & - & - & -3 & 1 & 1 \\ \cline{2-14}
		& $\Gamma_5^+$ & 6 & - & - & 0 & 0 & 0 & 6(-2) & - & - & 0 & 0 & 0 \\ \cline{2-14}
		& $\Gamma_5^-$ & 6 & - & - & 0 & 0 & 0 & -6(2) & - & - & 0 & 0 & 0 \\ \hline
	\end{tabular}
\end{table}
We follow the designation of point group irreps as in KDWS\cite{Koster_G:1963}. The real space lattice, Wyckoff positions, BZ, and RD  are indicated in Fig.\ref{fig:graphene}. The $\tilde{\rm G}^{\rm\bf k}$ at $\Gamma$, M, and K are D$_{6h}$, D$_{2h}$, and D$_{3h}$ respectively. The site symmetry groups of the Wyckoff poisitions 2b, 3d, and 1a are isomorphic to the point groups ${\rm D}_{3h}$, ${\rm D}_{2h}$ and ${\rm D}_{6h}$ respectively. The point group associated with this symmorphic layer group is ${\rm D}_{6h}$. It is important to know the correspondence of element/conjugacy class of these groups. While $\tilde{\rm G}^{\rm K}$ and $\tilde{\rm G}^{2b}$ are all isomorphic to the point group D$_{3h}$, they contain some distinct elements. Tab.\ref{tbl:grp_map} shows such correspondence between $\tilde{\rm G}^{\rm\bf k}$ and $\tilde{\rm G}^{\tau}$, and the point group F=$\tilde{\rm G}^{\Gamma}$. Specifically, the ${\rm p}_z$ orbital on Wyckoff position 2b transforms as $\Gamma_4$ without spin and $\Gamma_8$ with spin under $\tilde{\rm G}^{2b}$. The s orbital and p$_{xy}$ orbitals on Wyckoff position 2b transform as $\Gamma_1$ and $\Gamma_6$ of $\tilde{\rm G}^{2b}$ respectivelg..

Tab.\ref{tab:G80} shows evaluated value of $\chi_\phi$ in Eq.\eqref{eqn:band_decomp} for EBR centred on Wyckof position 2b and 3d. The value of $\chi_\phi$ is the same for all ${\rm\bf k}$ except at some HSPs on the surface of BZ where the gauge term modifies its value. This only occurs at K for EBR arising from Wyckoff position 2b and M for EBR arising from Wyckoff position 3d respectively. The modified value are given in bracket in the table. For EBR arising from Wyckoff position 1a, the value of $\chi_\phi$ is the same as that of character table from $\tilde{\rm G}^{1a}$ or D$_{6h}$ point group. The gauge term has no effect here due to absence of multiplicity. From these value of $\chi_\phi$, the decomposition of EBRs are calculated from Eq.\eqref{eqn:band_decomp} and tabulated in Tab.\ref{tab:spG80_decomp} for the HSPs in RD.
\begin{table}
	\caption{\label{tab:spG80_decomp}Decomposition of EBR induced from Wannier function centred on Wyckoff position 1a, 2b and 3d of honeycomb lattice with layer group 80 on restriction to $\mathcal{G}^{\rm\bf k}$ at HSPs in the BZ in terms of irreps of $\mathcal{G}^{\rm\bf k}$ induced from that of $\tilde{\rm G}^{\rm\bf k}$.} 	
	\begin{tabular}{|c|c|c|c|c|c|c|c|} \hline
	 \multicolumn{2}{|c|} {EBR} & \multirow{2}{*} {$\Gamma ({\rm D}_{6h})$} & \multirow{2}{*} {T (C$_{2v}$)} & \multirow{2}{*} {K (D$_{3h}$)} & \multirow{2}{*} {${\rm T}^\prime$(C$_{2v}$)} & \multirow{2}{*} {M(D$_{2h}$)} & \multirow{2}{*} {$\Sigma$(C$_{2v}$)} \\ \cline{1-2}
	 ${\bf \tau}_\alpha$ & $\mu$ of $\tilde{\rm G}^{\tau}$ & & & & & & \\ \hline
	 \multirow{8}{*}{1a} & $\Gamma_6^+$ & $\Gamma_6^+$ & ${\rm T}_1\oplus{\rm T}_2$ & ${\rm K}_6$ & ${\rm T}^\prime_1\oplus{\rm T}^\prime_2$ & ${\rm M}_1^+\oplus {\rm M}_3^+$ & $\Sigma_1\oplus\Sigma_4$\\ \cline{2-8} 
	 &  $\Gamma_5^-$ & $\Gamma_5^-$ & ${\rm T}_1\oplus{\rm T}_2$ & ${\rm K}_6$ & ${\rm T}^\prime_1\oplus{\rm T}^\prime_2$ & ${\rm M}_2^-\oplus{\rm M}_4^-$ & $\Sigma_1\oplus\Sigma_4$\\ \cline{2-8} 
	 &  $\Gamma_7^+$ & $\Gamma_7^+$ & ${\rm T}_5$ & ${\rm K}_7$ & ${\rm T}^\prime_5$ & ${\rm M}_5^+$ & $\Sigma_5$\\ \cline{2-8} 
	 & $\Gamma_8^+$ & $\Gamma_8^+$ & ${\rm T}_5$ & ${\rm K}_8$ & ${\rm T}^\prime_5$ & ${\rm M}_5^+$ & $\Sigma_5$\\ \cline{2-8} 
	 & $\Gamma_9^+$ & $\Gamma_9^+$ & ${\rm T}_5$ & ${\rm K}_9$ & ${\rm T}^\prime_5$ & ${\rm M}_5^+$ & $\Sigma_5$\\ \cline{2-8} 
	 & $\Gamma_7^-$ & $\Gamma_7^-$ & ${\rm T}_5$ & ${\rm K}_8$ & ${\rm T}^\prime_5$ & ${\rm M}_5^-$ & $\Sigma_5$\\ \cline{2-8} 
	 & $\Gamma_8^-$ & $\Gamma_8^-$ & ${\rm T}_5$ & ${\rm K}_7$ & ${\rm T}^\prime_5$ & ${\rm M}_5^-$ & $\Sigma_5$\\ \cline{2-8} 
	 & $\Gamma_9^-$ & $\Gamma_9^-$ & ${\rm T}_5$ & ${\rm K}_9$ & ${\rm T}^\prime_5$ & ${\rm M}_5^-$ & $\Sigma_5$\\ \hline 
	 \multirow{7}{*}{2b} & $\Gamma_4 ({\rm p}_z)$ & $\Gamma_4^+\oplus\Gamma_2^-$ & ${\rm T}_3\oplus {\rm T}_4$ & ${\rm K}_5$ & ${\rm T}^\prime_3\oplus {\rm T}^\prime_4$ & ${\rm M}_4^+\oplus{\rm M}_3^-$ & $2\Sigma_2$\\ \cline{2-8} 
	\multirow{2}{*}{} & \multirow{2}{*}{$\Gamma_5$} & \multirow{2}{*}{$\Gamma_5^+\oplus\Gamma_6^-$} & \multirow{2}{*}{2${\rm T}_3\oplus 2{\rm T}_4$} & \multirow{2}{*}{${\rm K}_3\oplus {\rm K}_4\oplus {\rm K}_5$} & \multirow{2}{*}{$2{\rm T}^\prime_3\oplus 2{\rm T}^\prime_4$} & ${\rm M}_2^+\oplus {\rm M}_4^+\oplus $ & \multirow{2}{*}{$2\Sigma_2\oplus 2\Sigma_3$}\\
	 & & & & & & ${\rm M}_1^-\oplus {\rm M}_3^-$ & \\ \cline {2-8}                                                                                                                                                                                                          
	 & $\Gamma_1 ({\rm s})$ & $\Gamma_1^+\oplus\Gamma_3^-$ & ${\rm T}_1\oplus {\rm T}_2$ & ${\rm K}_6$ & ${\rm T}^\prime_1\oplus {\rm T}^\prime_2$ & ${\rm M}_1^+\oplus {\rm M}_2^-$ & $2\Sigma_1$\\ \cline {2-8}                                                                                                                                                                                                          
	\multirow{2}{*}{} & \multirow{2}{*}{$\Gamma_6 ({\rm p}_{x,y})$} & \multirow{2}{*}{$\Gamma_6^+\oplus\Gamma_5^-$} & \multirow{2}{*}{2${\rm T}_1\oplus 2{\rm T}_2$} & \multirow{2}{*}{${\rm K}_1\oplus {\rm K}_2\oplus {\rm K}_6$} & \multirow{2}{*}{$2{\rm T}^\prime_1\oplus 2{\rm T}^\prime_2$} & ${\rm M}_1^+\oplus {\rm M}_3^+$ & \multirow{2}{*}{$2\Sigma_1\oplus 2\Sigma_4$}\\ 
	& & & & & & $\oplus {\rm M}_2^-\oplus {\rm M}_4^-$ & \\ \cline {2-8}                                                                                                                                                                                                          
	 & $\Gamma_7 ({\rm s})$ & $\Gamma_7^+\oplus\Gamma_8^-$ & $2{\rm T}_5$ & ${\rm K}_8\oplus {\rm K}_9$ & $2{\rm T}^\prime_5$ & ${\rm M}_5^+\oplus {\rm M}_5^-$ &$2\Sigma_5$  \\ \cline{2-8}
	 & $\Gamma_8 ({\rm p}_z)$ & $\Gamma_8^+\oplus\Gamma_7^-$ & $2{\rm T}_5$ & ${\rm K}_7\oplus {\rm K}_9$ & $2{\rm T}^\prime_5$ & ${\rm M}_5^+\oplus {\rm M}_5^-$ & $2\Sigma_5$\\ \cline{2-8}
	 & $\Gamma_9$ & $\Gamma_9^+\oplus\Gamma_9^-$ & $2{\rm T}_5$ & ${\rm K}_7\oplus {\rm K}_8$ & $2{\rm T}^\prime_5$ & ${\rm M}_5^+\oplus {\rm M}_5^-$ & $2\Sigma_5$ \\ \hline
	 	 \multirow{6}{*}{3d} & $\Gamma_1^+$ & $\Gamma_1^+\oplus\Gamma_6^+$ & $2{\rm T}_1\oplus {\rm T}_2$ & ${\rm K}_1\oplus {\rm K}_6$ & $2{\rm T}^\prime_1\oplus {\rm T}^\prime_2$ & ${\rm M}_1^+\oplus{\rm M}_2^-\oplus{\rm M}_4^-$ & $2\Sigma_1\oplus\Sigma_4$\\ \cline{2-8} 
	 & $\Gamma_2^-$ & $\Gamma_3^-\oplus\Gamma_5^-$ & ${\rm T}_1\oplus 2{\rm T}_2$ & ${\rm K}_2\oplus {\rm K}_6$ & ${\rm T}^\prime_1\oplus 2{\rm T}^\prime_2$ & ${\rm M}_1^+\oplus{\rm M}_3^+\oplus{\rm M}_2^-$ & $2\Sigma_1\oplus \Sigma_4$\\  \cline{2-8}
	 & $\Gamma_4^+$ & $\Gamma_4^+\oplus\Gamma_5^+$ & $2{\rm T}_3\oplus {\rm T}_4$ & ${\rm K}_3\oplus {\rm K}_5$ & $2{\rm T}^\prime_3\oplus {\rm T}^\prime_4$ & ${\rm M}_4^+\oplus{\rm M}_1^-\oplus{\rm M}_3^-$ & $2\Sigma_2\oplus \Sigma_3$\\  \cline{2-8}
	 & $\Gamma_3^-$ & $\Gamma_2^-\oplus\Gamma_6^-$ & ${\rm T}_3\oplus 2{\rm T}_4$ & ${\rm K}_4\oplus {\rm K}_5$ & ${\rm T}^\prime_3\oplus 2{\rm T}^\prime_4$ & ${\rm M}_2^+\oplus{\rm M}_4^+\oplus{\rm M}_3^-$ & $2\Sigma_2\oplus \Sigma_3$\\  \cline{2-8}
	 & $\Gamma_5^+$ & $\Gamma_7^+\oplus\Gamma_8^+\oplus\Gamma_9^+$ & $3{\rm T}_5$ & ${\rm K}_7\oplus{\rm K}_8\oplus {\rm K}_9$ & $3{\rm T}^\prime_5$ & ${\rm M}_5^+\oplus2{\rm M}_5^-$ & $3\Sigma_5$\\  \cline{2-8}
	 & $\Gamma_5^-$ & $\Gamma_7^-\oplus\Gamma_8^-\oplus\Gamma_9^-$ & $3{\rm T}_5$ & ${\rm K}_7\oplus{\rm K}_8\oplus {\rm K}_9$ & $3{\rm T}^\prime_5$ & $2{\rm M}_5^+\oplus{\rm M}_5^-$ & $3\Sigma_5$\\ \hline 
	\end{tabular}
\end{table}

\begin{table}[t]
\caption{\label{tab:IBR}Decomposition of IBRs arising from some EBRs on Wyckoff position 2b and 3d. $\mathcal{G}=80(L)$. For type 2 IBRs, the equivalent EBRs are identified in the last column.}
\begin{tabular}{c|c|c|c|c|c}
EBR & IBR Type & $\Gamma$ & M & K & Equiv. EBR on 1a\\ \hline
\parbox[t]{2mm}{\multirow{3}{*}{\rotatebox[origin=c]{90}{$\Gamma_1$ on 2b}}} & Type 1 & $\Gamma_1^+\oplus\Gamma_3^-$ & M$_1^+\oplus$M$_2^-$ & K$_6$\\ \cline{2-6}
& Type 2$\beta$ & $\Gamma_1^+$ & M$_1^+$ & K$_1$ & $\Gamma_1^+$ \\ \cline{2-6}
& Type 2$\beta$ & $\Gamma_3^-$ & M$_2^-$ & K$_2$ & $\Gamma_3^-$ \\ \hline
\parbox[t]{2mm}{\multirow{3}{*}{\rotatebox[origin=c]{90}{$\Gamma_4$ on 2b}}} & Type 1 & $\Gamma_4^+\oplus\Gamma_2^-$ & M$_3^+\oplus$M$_3^-$ & K$_5$\\ \cline{2-6}
& Type 2$\beta$ & $\Gamma_4^+$ & M$_3^+$ & K$_3$ & $\Gamma_4^+$\\ \cline{2-6}
& Type 2$\beta$ & $\Gamma_2^-$ & M$_4^-$ & K$_4$ & $\Gamma_2^-$\\ \hline
\parbox[t]{2mm}{\multirow{5}{*}{\rotatebox[origin=c]{90}{$\Gamma_6$ on 2b}}} & Type 1 & $\Gamma_6^+\oplus\Gamma_5^-$ & M$_1^+\oplus$M$_3^+\oplus$M$_2^-\oplus$M$_4^-$ & K$_1\oplus$K$_2\oplus$K$_6$ & \\ \cline{2-6}
& Type 2$\alpha$ & $\Gamma_6^+$ & M$_1^+\oplus$M$_3^+$ & K$_6$ & $\Gamma_6^+$ \\ \cline{2-6}
& Type 3 & $\Gamma_5^-$ & M$_2^-\oplus$M$_4^-$ & K$_1\oplus$K$_2$ & \\ \cline{2-6}
& Type 2$\alpha$ & $\Gamma_5^-$ & M$_2^-\oplus$M$_4^-$ & K$_6$ & $\Gamma_5^-$\\ \cline{2-6}
& Type 3 & $\Gamma_6^+$ & M$_1^+\oplus$M$_3^+$ & K$_1\oplus$K$_2$ & \\ \hline
\parbox[t]{2mm}{\multirow{3}{*}{\rotatebox[origin=c]{90}{$\Gamma_8$ on 2b}}} & Type 1 & $\Gamma_8^+\oplus\Gamma_7^-$ & M$_5^+\oplus$M$_5^-$ & K$_7\oplus$K$_9$ & \\ \cline{2-6}
& Type 2$\beta$ & $\Gamma_8^+$ & M$_5^+$ & K$_8$ & $\Gamma_8^+$ \\ \cline{2-6}
& Type 2$\beta$ & $\Gamma_7^-$ & M$_5^-$ & K$_8$ & $\Gamma_7^-$ \\ \hline
\parbox[t]{2mm}{\multirow{4}{*}{\rotatebox[origin=c]{90}{$\Gamma_1^+$ on 3d}}} & Type 1 & $\Gamma_1^+\oplus\Gamma_6^+$ & M$_1^+\oplus$M$_2^-\oplus$M$_4^-$ & K$_1\oplus$K$_6$\\ \cline{2-6}
& Type 2$\alpha$ & $\Gamma_1^+$ & M$_1^+$ & K$_1$ & $\Gamma_1^+$ \\ \cline{2-6}
& Type 3 & $\Gamma_6^+$ & M$_2^-\oplus$M$_4^-$ & K$_6$ & \\ \cline{2-6}
& Type 2$\beta$ & $\Gamma_6^+$ & M$_1^+\oplus$M$_3^+$ & K$_6$ & $\Gamma_6^+$\\ \hline
\end{tabular}
\end{table}

As indicated in the main text, IBRs, arising from linear combination of EBR components, are important concepts in the study of topology of the bands. Like EBRs, they are identified by the connectivity and their decomposition at HSPs in terms of irreps of $\mathcal{G}^{\rm\bf k}$. Tab.\ref{tab:IBR} shows IBR decompositions and type arising from some EBRs centred on Wyckoff position 2b and 3d.

Use of Zak basis present some issue relating to explicit ${\rm\bf k}$ dependence of its representation matrices . However, Zak and TB basis span the same vector space for a given ${\rm\bf k}$ and they return the same decompositions. At HSPs on the surface of BZ in particular, the impact of the gauge term is the same on both. 

\section{Group theoretical basis of tight binding theory}
\label{sec:III}
In TB theory, the basis used to express the eigen functions are the Bloch sums shown in Eq.\eqref{eqn:TB_basis} or \eqref{eqn:Zak_basis}. Yet the evaluation of the TB model are performed via sequential sums over `atoms' (Wyckoff positions) on shells of the same radius. This clearly indicate that the translational symmetry must be factored out in the evaluation of the matrix element. The aim of this appendix is to first derive the general form of the Hamiltonian in ${\rm\bf k}$ space and then derive the specific form of the Hamiltonian at a given ${\rm\bf k}\in$RD and neighbouring distance using Wigner-Eckart/general matrix elements theorem/method of invariants\cite{Eckart_C:1930, Koster_G:1958, Bir_G_L:1974} under the TB approximation.
\subsection{General form of matrix representation of $\hat{H}$ with respect to EBRs}
First of all, basis of representations of space group $\mathcal{G}$ derived from any BR but at {\em different} wave vectors are orthogonal,  i.e.
\[
\left<{\rm\bf k}^\prime, \bm{\sigma}_\alpha, \mu, i\mid{\rm\bf k}, \bm{\tau}_\beta, \nu, j\right>=0, \quad\mbox{unless~~}{\rm\bf k}^\prime={\rm\bf k}+{\rm\bf g}.
\]
Since  $\mathcal{G}$ is the symmetry group of the Hamiltonian operator $\hat{H}$, its matrix representation with respect to such basis are zero unless the wave vectors of the bra and ket are the same. From group rearrangement theorem, its matrix representation are given by
\begin{eqnarray}
\left<{\rm\bf k}, \bm{\sigma}_\alpha, \mu, i\mid\hat{H}
\mid{\rm\bf k}, \bm{\tau}_\beta, \nu, j\right>&=&\frac{1}{|\mathcal{G}|}\sum_{g\in\mathcal{G}} g\bullet\left<{\rm\bf k}, \bm{\sigma}_\alpha, \mu, i\mid\hat{H}
\mid{\rm\bf k}, \bm{\tau}_\beta, \nu, j\right>\notag \\
&=&\frac{1}{|\mathcal{G}|}\sum_{g\in\mathcal{G}} \left<{\rm\bf k}, \bm{\sigma}_\alpha, \mu, i\mid\hat{S}(g)^\dag \underbrace{\hat{S}(g)\hat{H} \hat{S}(g)^\dag}_{=\hat{H}} \hat{S}(g)
\mid {\rm\bf k}, \bm{\tau}_\beta, \nu, j\right>\notag\\
&=&\frac{1}{|\mathcal{G}|}\sum_{g\in\mathcal{G}} \left<{\rm\bf k}, \bm{\sigma}_\alpha, \mu, i\mid\hat{S}(g)^\dag \hat{H}  \hat{S}(g)
\mid {\rm\bf k}, \bm{\tau}_\beta, \nu, j\right>\label {eqn:grp_rearrange}
\end{eqnarray}
which refers to the full matrix representation with row and column indexed by distinct ${\rm\bf k}\in\{\ast{\rm\bf k}\}, \bm{\sigma}_\alpha, \mu, i$ labels. The multiplicity of the of $\bm{\sigma}_\alpha$ and $\bm{\tau}_\beta$ are $m$ and $n$ respectively. A quick conclusion can be made from Eq.\eqref{eqn:grp_rearrange}. If we act on both side by an element $\{R|{\rm\bf v+t}\}$ which permute ${\rm\bf k}$ index among $\{\ast {\rm\bf k}\}$, the ${\rm\bf k}$ index on the LHS would change, yet the RHS remains unchanged due to group rearrangement theorem. Hence each diagonal sub block indexed by ${\rm\bf k}^\prime\in\{\ast {\rm\bf k}\}$ have the same functional dependence on ${\rm\bf k}^\prime$ and sub-block form. The off-diagonal blocks must be zero. i.e. the matrix representation of $\hat{H}$ with respect to EBR basis are in block diagonal form with each block indexed by ${\rm\bf k}^\prime\in\{\ast {\rm\bf k}\}$. For the {\rm blocks} corresponding to arms ${\rm\bf k}$ and $R{\rm\bf k}$ of $\{\ast {\rm\bf k}\}$:
\begin{equation}
S^\dag(\{R|{\rm\bf v+t}\})H(R{\rm\bf k})S(\{R|{\rm\bf v+t}\})=H({\rm\bf k}). \label{eqn:k_block}
\end{equation}
Thus the symmetry group of {\em a sub-block} $H({\rm\bf k})$ is $\mathcal{G}^{\rm\bf k}$. Any other element of $\{R|{\rm\bf v+t}\}\in(\mathcal{G}\setminus\mathcal{G}^{\rm\bf k})$ transform the $H({\rm\bf k})$ to other block indexed by $H(R{\rm\bf k})$ according to Eq.\eqref{eqn:k_block}.

\subsection{Form of matrix representation of $\hat{H}$ under TB approximation}
Let's define
\[
\left<{\rm\bf r}\mid({\rm\bf t}_\ell+\bm{\tau}_\alpha),\mu,i\right>=W_{\bm{\tau}_\alpha}^{\mu,i}({\rm\bf r}-({\rm\bf t}_\ell+\bm{\tau}_\alpha)).
\]
where the Wainner functions are labeled by its localisation centre $({\rm\bf t}_\ell+\bm{\tau}_\alpha)$, irrep label $\mu$ and its index $i$. The transformation properties of a Wannier function centred at ${\rm\bf t}_\ell+{\bm \tau}_\alpha$ are given by Eq.\eqref{eqn:wannier_tfmb} where
\[
{\rm\bf t}_{\ell^\prime}+{\bm \tau}_\beta=\{R\mid{\rm\bf v+t}\}({\rm\bf t}_{\ell}+{\bm \tau}_\alpha)
\]
Hence the matrix representation of $\hat{H}$ with respect to the EBR basis can be expanded as
\begin{align}
H({\rm\bf k})&=\left<{\rm\bf k}, \bm{\sigma}_\alpha,\mu,i\mid\hat{H}\mid{\rm\bf k}, \bm{\tau}_\beta, \nu,j\right> \notag\\
&=\sum_\ell\sum_{\ell^\prime}\exp\{-i{\rm\bf k}\cdot({\rm\bf t}_\ell+\bm{\sigma}_\alpha)\left<({\rm\bf t}_\ell+\bm{\sigma}_\alpha),\mu,i\mid\hat{H}\mid({\rm\bf t}_{\ell^\prime}+\bm{\tau}_\beta),\nu,j\right>\exp\{i{\rm\bf k}\cdot({\rm\bf t}_{\ell^\prime}+\bm{\tau}_\beta)\notag\\
&=\sum_\ell\sum_{\ell^\prime}\exp\{i{\rm\bf k}\cdot[({\rm\bf t}_{\ell^\prime}+\bm{\tau}_\beta)-({\rm\bf t}_\ell+\bm{\sigma}_\alpha)]\}\left<({\rm\bf t}_\ell+\bm{\sigma}_\alpha),\mu,i\mid\hat{H}\mid({\rm\bf t}_{\ell^\prime}+\bm{\tau}_\beta),\nu,j\right>
\label{eqn:Hmatrix}
\end{align}
where the phase factor are defined by the difference in localisation centre of the ket and bra. For Eq.\eqref{eqn:Hmatrix} to be invariant under $\mathcal{G}$, the action of $g\in\mathcal{G}$ can only transform (permute) the terms in the summation. Consider the transformation properties of the general term in the summations. Under the action of an element of space group $\{R\mid{\rm\bf v+t}\}$, the Wannier functions transforms as Eq.\eqref{eqn:wannier_tfmb}, but the phase term is a constant independent of ${\rm\bf r}$. We need to introduce the action of $\{R\mid{\rm\bf v+t}\}$ on the localisation centres on the phase term in order to keep such a transformed general term within $H({\rm\bf k})$ in Eq.\eqref{eqn:Hmatrix}. 
\[
\begin{aligned}
\{R\mid{\rm\bf v+t}\}\bullet&\left[\exp\{i{\rm\bf k}\cdot[({\rm\bf t}_{\ell^\prime}+\tau_\beta)-({\rm\bf t}_\ell+\sigma_\alpha)]\}\left<{\rm\bf t}_\ell,\sigma_\alpha,\mu,i\mid\hat{H}\mid{\rm\bf t}_{\ell^\prime},\tau_\beta,\nu,j\right>\right]\\
&=\exp\{i{\rm\bf k}\cdot\underbrace{\{R\mid{\rm\bf v+t}\}\bullet [({\rm\bf t}_{\ell^\prime}+\tau_\beta)-({\rm\bf t}_\ell+\sigma_\alpha)]}\} \\
&\quad\quad\sum_{\gamma,p}D^{\bm{\sigma},\mu}(R)^{\ast}_{\bm{\sigma}_\gamma,p; \bm{\sigma}_\alpha,i}\sum_{\delta,q}D^{\bm{\tau},\nu}(R)_{\bm{\tau}_\delta,q; \bm{\tau}_\beta,j}\left<{\rm\bf t}_{\ell^{\prime\prime}},\sigma_\gamma,\mu,p\mid\hat{H}\mid{\rm\bf t}_{\ell^{\prime\prime\prime}},\tau_\delta,\nu,q\right>
\end{aligned}
\]
where $\{R\mid{\rm\bf v+t}\}({\rm\bf t}_\ell+\sigma_\alpha)=({\rm\bf t}_{\ell^{\prime\prime}}+\sigma_{\gamma^\prime})$ and $\{R\mid{\rm\bf v+t}\}({\rm\bf t}_{\ell^\prime}+\tau_\beta)=({\rm\bf t}_{\ell^{\prime\prime\prime}}+\tau_{\delta^\prime})$. The permutational nature of $D^{\bm{\tau},\nu}(R)$ ensures only $\tau_{\delta^\prime}$ and $\sigma_{\gamma^\prime}$ terms surrive. In the analysis here, we keep ${\rm\bf k}$ unchanged in contrast to derivation of Eq.\eqref{eqn:tfmTB}. Since $\{R\mid{\rm\bf v+t}\}$ is distance preserving,
\[
\begin{aligned}
({\rm\bf t}_{\ell^{\prime\prime\prime}}+\tau_{\delta^\prime})-({\rm\bf t}_{\ell^{\prime\prime}}+\sigma_{\gamma^\prime})&={\boldsymbol\rho}^d_r=\{R\mid{\rm\bf v+t}\}[({\rm\bf t}_{\ell^\prime}+\tau_\beta)-({\rm\bf t}_\ell+\sigma_\alpha)]\\
\left|({\rm\bf t}_{\ell^{\prime\prime\prime}}+\tau_{\delta^\prime})-({\rm\bf t}_{\ell^{\prime\prime}}+\sigma_{\gamma^\prime})\right|&=\rho^d=\left|({\rm\bf t}_{\ell^\prime}+\tau_\beta)-({\rm\bf t}_\ell+\sigma_\alpha)\right|
\end{aligned}
\]
Translation has no impact on distances between two points. Thus for a given $({\rm\bf t}_\ell+\sigma_\alpha)$ and $({\rm\bf t}_{\ell^\prime}+\tau_\beta)$, the action of all $\{R\mid{\rm\bf v+t}\}\in\mathcal{G}$ generate a finite set of vectors (vectors with different $\sigma_\gamma$ but the same direction/magnitude are considered distinct). 
\[
\boldsymbol\rho^d_r=({\rm\bf t}_{\ell^{\prime\prime\prime}}+\tau_{\delta^\prime})-({\rm\bf t}_{\ell^{\prime\prime}}+\sigma_{\gamma^\prime})=R\bullet[({\rm\bf t}_{\ell^\prime}+\tau_\beta)-({\rm\bf t}_\ell+\sigma_\alpha)]
\]
with the same magnitude $\rho^d$. The action of $\{R\mid{\rm\bf v+t}\}$ permutes the set of vectors $\boldsymbol\rho^d_r$ and we can express this as a permutation matrix
\[
\boldsymbol\rho_r^d=\sum_sD(R)^\rho_{sr}\boldsymbol\rho_s.\quad\mbox{or}\quad\exp\{i{\rm\bf k}\cdot\boldsymbol\rho^d_r\}=\sum_sD(R)^\rho_{sr}\exp\{i{\rm\bf k}\cdot\boldsymbol\rho^d_s\}
\]
Thus the transformation properties of a general term in the matrix representation of $\hat{H}$ is given by
\begin{align}
\{R\mid{\rm\bf v+t}\}\bullet&\left[\exp\{i{\rm\bf k}\cdot\boldsymbol\rho^d_1\}\left<{\rm\bf t}_\ell,\sigma_\alpha,\mu,i\mid\hat{H}\mid{\rm\bf t}_{\ell^\prime},\tau_\beta,\nu,j\right>\right]\notag\\
&=\sum_kD^\rho(R)_{kr}\sum_{\gamma,p}D^{\bm{\sigma},\mu}(R)^{\ast}_{\bm{\sigma}_\gamma,p; \bm{\sigma}_\alpha,i}\sum_{\delta,q}D^{\bm{\tau},\nu}(R)_{\bm{\tau}_\delta,q; \bm{\tau}_\beta,j}\notag\\
&\quad\left[\exp\{i{\rm\bf k}\cdot\boldsymbol\rho^d_k\}\left<{\rm\bf t}_{\ell^{\prime\prime}},\sigma_\gamma,\mu,p\mid\hat{H}\mid{\rm\bf t}_{\ell^{\prime\prime\prime}},\tau_\delta,\nu,q\right>\right].
\label{eqn:HTerm}
\end{align}
i.e. $\{R\mid{\rm\bf v+t}\}$ generate another term in the summation in Eq.\eqref{eqn:Hmatrix} from the starting term, subject to the constraint on the magnitude of $\boldsymbol\rho^d_r$. Now consider the summation
\begin{equation}
H^{\{d\}}({\rm\bf k})=\frac{1}{M^d}\sum_{g\in\mathcal{G}}g\bullet\left[\exp\{i{\rm\bf k}\cdot\boldsymbol\rho^d_1\}\left<{\rm\bf t}_\ell,\sigma_\alpha,\mu,i\mid\hat{H}\mid{\rm\bf t}_{\ell^\prime},\tau_\beta,\nu,j\right>\right],\label{eqn:Hgroupsum}
\end{equation}
where $M^d$ is the ratio of the order of point group F to the dimension of $D^\rho(R)$. With a view of decomposition of $\mathcal{G}$ in Eq.\eqref{eqn:A1}, the sum can be considered as over the crystallographic orbits of $\sigma_\alpha$ in the bra (generated by $\mathcal{T}$) and all permitted $\alpha$ (generated by the coset representative) but with a constraint of separation of the localisation centres of the two Wannier functions in the ket and bra equal to $\rho^d$. In fact, the sum in Eq.\eqref{eqn:Hgroupsum} reproduces the relevant terms in Eq.\ref{eqn:Hmatrix} multiple times as defined by $M_d$, depending on the constraining radius $\rho^d$. It is not difficult to see that this sum is invariant due to group rearrangement theorem. To complete the sum over the crystallgraphic orbit in $\bm\tau_\beta$ in the ket, we need to consider all possible radius $\rho^d$ as in:
\begin{equation}
H({\rm\bf k})=H^{ZNN}({\rm\bf k})+H^{FNN}({\rm\bf k})+H^{SNN}({\rm\bf k})+\cdots+H^{\{d\}}({\rm\bf k})+\cdots, \label{eqn:TB_H}
\end{equation}
where ZNN, FNN, SNN stands for `on site', `first nearest neighbour' and `second nearest neighbour' etc as index d. The degree of approximation is set by the truncation of the series with the understanding that the `reduced tensor element' decreases rapidly with increasing $\rho^d$.

Given the transformation properties of a term in $H^{\{d\}}({\rm\bf k})_{\sigma_\alpha,\mu,i; \tau_\beta,\nu,j}$ in Eq.\eqref{eqn:HTerm}, the form of $H^{\{d\}}({\rm\bf k})$ can be obtained using general matrix element theorem\cite{Koster_G:1958} by constructing projection operator of the trivial representation (See Appx.\ref{sec:general_matrix} and \ref{sec:TB_Ham}). There is no dependence on the lattice translation ${\rm\bf t}$ or non-lattice translation ${\rm\bf v}$. One only needs to sum over the point group elements for such projection operator. It is important to note that there are linearly independent forms under a particular $\rho^d$ with corresponding hopping parameters (e.g. the second nearest neighbour $H^{SNN}({\rm\bf k})$ in the example in next section). The TB approximation asserts that the `reduced tensor element' or hopping parameter of $H^{\{d\}}({\rm\bf k})_{\sigma_\alpha,\mu,i; \tau_\beta,\nu,j}$ is dependent only on the distance $\rho^d=\left|({\rm\bf t}_{\ell^{\prime\prime\prime}}+\tau_{\delta^\prime})-({\rm\bf t}_{\ell^{\prime\prime}}+\sigma_{\gamma^\prime})\right|=\left|\rho^d_1\right|=\left|({\rm\bf t}_{\ell^\prime}+\tau_\beta)-({\rm\bf t}_\ell+\sigma_\alpha)\right|$. The orientation dependence on such vectors are given by the `angular dependence' on the EBR index $\tau_\beta,\nu,j$. Implicit in this statement is that the contribution from all pair of Wannier functions, subject to the constraint $\left|({\rm\bf t}_{\ell^{\prime\prime\prime}}+\tau_{\delta^\prime})-({\rm\bf t}_{\ell^{\prime\prime}}+\sigma_{\gamma^\prime})\right|=\left|\rho^d\right|$, to the matrix representation of $\hat{H}$ form a representation of $\mathcal{G}$. The general matrix element theorem can then be applied to these contributions to $H^{\{d\}}({\rm\bf k})$. The results apply to all diagonal blocks indexed by ${\rm\bf k}^\prime\in\{\ast \rm\bf k\}$.

\section{General matrix element theorem and method of invariant\label{sec:general_matrix}}
This section provides a brief description on how to obtain the angular dependent part of general matrix element theorem\cite{Koster_G:1958} or method of invariant\cite{Luttinger_J_M:1956, Bir_G_L:1974}, utilising projection operator technique. This is a generalisation of the Wigner Eckart theorem\cite{Eckart_C:1930}.

Consider a physically measurable quantity described by some matrix element of some vector (tensor) operator $A^\omega_p$ between some eigenstate $\left|\psi^\nu_j\right>$ where $\mu, \omega, \nu$ are irrep labels of the symmetry group of the Hamiltonian G. 
\begin{equation}
A^{\omega, p}_{\mu i, \nu j}=\left<\psi^\mu_i\mid A^{\omega}_p\mid\psi^\nu_j\right>.
\end{equation}
where each of the components transform under symmetry operator $R$ according to 
\begin{subequations}
\begin{align}
{A^\omega_p}^\prime&=\sum_qD^{\omega}(R)_{qp}A^\omega_q \\
\left|\psi^\nu_j\right>^\prime&=\sum_rD^{\nu}(R)_{rj}\left|\psi^\nu_r\right> \\
\left<\psi^\mu_i\right|^\prime&=\sum_sD^{\mu}(R)^\ast_{si}\left<\psi^\mu_s\right|.
\end{align}
\end{subequations}
Therefore the whole matrix element transform under the action of $R\in {\rm G}$ according to
\[
D(R)=D^{\omega}(R)\otimes D^{\mu}(R)^\ast\otimes D^{\nu}(R),
\]
with dictionary order of $\omega,p; \mu, i; \nu, j$ as the composite row and column index of $D(R)$. Since the quantities are invariant under the action of R, we have
\[
A^{\omega, p}_{\mu i, \nu j}=\frac{1}{|{\rm G}|}\sum_{g\in{\rm G}}g\bullet A^{\omega, p}_{\mu i, \nu j}=\underbrace{\frac{1}{|{\rm G}|}\sum_R D(R)}_{\mbox{Projection operator of }\Gamma_1}A^{\omega, p^\prime}_{\mu i^\prime, \nu j^\prime}
\]
Thus the angular dependent part can be projected out with the projection operator of the trivial representation and trial column vector. The multiple allowed linearly independent angular dependent parts with corresponding reduced tensor elements are obtained from linearly independent results of projection.

The requirement of $\omega, \mu,$ and $\nu$ to be irreps is not essential. Provided the entities $A^\omega_p$, $\left<\psi^\mu_i\right|$ and $\left|\psi^\nu_j\right>$ are basis of representations of the group (direct sum of irreps), the methodology still applies. In the particular example below, $D^{\rho}(R)$ is not necessarily irreducible and the basis of Hamiltonian are normally CBRs.

\section{Application of method of invariant in obtaining TB Hamiltonian}
\label{sec:TB_Ham}
This section illustrate how to use the general matrix element theorem\cite{Koster_G:1958} or method of invariant\cite{Bir_G_L:1974, Luttinger_J_M:1956} to obtain $H({\rm\bf k})^{\{d\}}$ for single layer graphene. The procedure here evaluate the first three terms in Eq.\eqref{eqn:TB_H} for the EBR basis induced from spin-less and spin-full p$_z$ orbitals centred on Wyckoff position 2b.

The space group of single layer graphene crystal is the sub-periodic layer group 80 (P6/mmm). The point group of the graphene lattice is ${\rm D}_{6h}$, The lattice is shown in Fig.\ref{fig:graphene}. The carbon atoms are located on the Wyckoff position $\bm{\tau}={\rm 2b}$. We label the two equivalent Wyckoff positions as $\bm{\tau}_a$ and $\bm{\tau}_b$. The site symmetry group is isomorphic to ${\rm D}_{3h}$. The coset representatives in the decomposition in Eq.\eqref{eqn:A4} are chosen as $R^{\bm{\tau}}_a=E$ and $R^{\bm{\tau}}_b=I$ for $\bm{\tau}_a$ and $\bm{\tau}_b$ respectively. Hence the ${\rm p}_z$ orbitals on these atomic sites form the $\Gamma_4$ or $\Gamma_8$ irrep of the site symmetry group $\tilde{\rm G}^{\bm{\tau}}={\rm D}_{3h}$ for single group or double group consideration. 

We are concerned with the interaction between orbitals located on the equivalent Wyckoff positions. Since the multiplicity of $\bm{\tau}$ is 2, the site representation $D^{\bm{\tau}}(R)$ is a $2\times 2$ matrix and given by
\[
D^{\bm{\tau}}(R)=\left\{\begin{array}{ll}
\begin{pmatrix} 1 & 0 \\ 0 & 1\end{pmatrix}\quad&\mbox{if~}R\in {\rm D}_{3h} (\tilde{\rm G}^{2b})\\
\begin{pmatrix} 0 & 1 \\ 1 & 0\end{pmatrix}\quad&\mbox{if~}R\notin {\rm D}_{3h} (\tilde{\rm G}^{2b})\\	
\end{array}\right.
\]
As $D^{\bm{\tau}}(R)$ represents a permutation of the equivalent Wyckoff positions, there is only one unity entry in each row or column with others equal to 0. Let $D^{\bm{\tau}}(R)_{\alpha,\beta}=1$, then the mapping in Eq.\eqref{eqn:A7} yield
\[
R^{\bm{\tau}}_\leftarrow(R, R^{\bm{\tau}}_\alpha)= {R^{\bm{\tau}}_\alpha}^{-1}R R^{\bm{\tau}}_\beta
\]
In this particular case, the mapping is particularly simple and given by
\[
R^{\bm{\tau}}_\leftarrow(R, R^{\bm{\tau}}_\alpha)=\left\{\begin{array}{ll}
R\quad&\mbox{if~}R\in {\rm D}_{3h} (\tilde{\rm G}^{2b})\\
RI\quad&\mbox{if~}R\notin {\rm D}_{3h} (\tilde{\rm G}^{2b})	
\end{array}\right.
\]  
Then the representation matrices 
\[
D^{\bm{\tau},\nu}(R)=D^{\bm{\tau}}(R)\otimes D^{\nu}(R^{\bm{\tau}}_\leftarrow(R, R^{\bm{\tau}}_\alpha)),\quad\quad\quad\forall R\in {\rm F}.
\]
can be constructed. It is important to recognise that the mapping from $R$ to $R^{\bm{\tau}}_\leftarrow(R, R^{\bm{\tau}}_\alpha)$ in the second matrix function is dependent on the row index of $D^{\bm{\tau}}(R)$. The same approach is required to construct $D^{\bm{\sigma},\mu}(R)$. In this particular case, it is the same as $D^{\bm{\tau},\nu}(R)$.

\begin{figure}[htbp]
\includegraphics[width=0.9\textwidth]{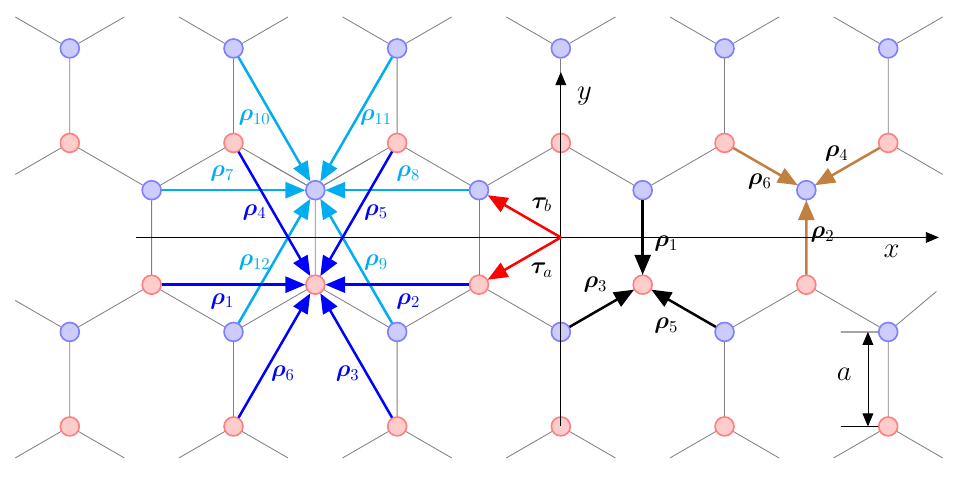}
\caption{Distinct radial vectors produced by symmetry operation for first neares neighbours (black \& brown) and second nearest neighbours (blue \& cyan).}
\label{fig:graphenea}
\end{figure}

We are now in a position to consider a particular choice of ${\rm\bf t}_\ell+\bm{\sigma}_\alpha$ and ${\rm\bf t}_{\ell^\prime}+\bm{\tau}_\beta$ (or $\rho^d$) in the partition of sum over all Wannier functions. The relevant radial vectors are shown in Fig.\ref{fig:graphenea} for the first nearest neighbour (FNN) and second nearest neighbour (SNN) sets.

\underline{Onsite interaction:} ${\rm\bf t}_\ell+\bm{\sigma}_\alpha={\rm\bf t}_{\ell^\prime}+\bm{\tau}_\beta=\bm{\tau}_a;\;\;\bm{\rho}_1=\bm{0}.$

Hence the phase factor $\exp[i{\rm\bf k}\cdot\bm{\rho}]=1$. Therefore $D^{\bm{\rho}}(R)=1$. (Strictly speaking, there is a $\boldsymbol\rho_2=\bm{0}$ corresponding to hopping from $\bm\tau_b$ to $\bm\tau_b$. We would need formally to treat $D^{\rho}(R)$ as $2\times 2$ matrix. But taking on one phase term of '1' does not alter the result.) We can then construct the projection operator
\[
\mathcal{P}^{\Gamma_1^+}=\frac{1}{|{\rm D}_{6h}|}\sum_{R\in{\rm D}_{6h}} D^{\bm{\rho}}(R)\otimes D^{\bm{\sigma},\mu}(R)^\ast\otimes D^{\bm{\tau},\nu}(R)
\]
For single group, setting $\bm{\sigma}=\bm{\tau}, \mu=\nu=\Gamma_4$, this yield
\[
\mathcal{P}^{\Gamma_1^+}=\frac{1}{2}\begin{pmatrix} 1 & 0 & 0 & 1 \\  0 & 1 & 1 & 0 \\ 0 & 1 & 1 & 0 \\  1 & 0 & 0 & 1  	\end{pmatrix}
\]
Projecting from $[1, 0, 0, 0]^T$ (this correspond to the interaction that specified this shell), we obtain:
\[
H=E^{\bm{\tau}}\begin{pmatrix} 1 & 0 \\ 0 & 1\end{pmatrix}
\]
where $E^{\bm{\tau}}$ is the material dependent parameter.  

For double group, setting $\bm{\sigma}=\bm{\tau}=\bm{\tau}_a, \mu=\nu=\Gamma_8$, this yield
\[
\mathcal{P}^{\Gamma_1^+}=\frac{1}{8}\left(\begin{array}{cccccccccccccccc} 1 & 0 & 0 & 0 & 0 & 1 & 0 & 0 & 0 & 0 & 1 & 0 & 0 & 0 & 0 & 1\\ 0 & 0 & 0 & 0 & 0 & 0 & 0 & 0 & 0 & 0 & 0 & 0 & 0 & 0 & 0 & 0\\ 0 & 0 & 1 & 0 & 0 & 0 & 0 & 1 & 1 & 0 & 0 & 0 & 0 & 1 & 0 & 0\\ 0 & 0 & 0 & 0 & 0 & 0 & 0 & 0 & 0 & 0 & 0 & 0 & 0 & 0 & 0 & 0\\ 0 & 0 & 0 & 0 & 0 & 0 & 0 & 0 & 0 & 0 & 0 & 0 & 0 & 0 & 0 & 0\\ 1 & 0 & 0 & 0 & 0 & 1 & 0 & 0 & 0 & 0 & 1 & 0 & 0 & 0 & 0 & 1\\ 0 & 0 & 0 & 0 & 0 & 0 & 0 & 0 & 0 & 0 & 0 & 0 & 0 & 0 & 0 & 0\\ 0 & 0 & 1 & 0 & 0 & 0 & 0 & 1 & 1 & 0 & 0 & 0 & 0 & 1 & 0 & 0\\ 0 & 0 & 1 & 0 & 0 & 0 & 0 & 1 & 1 & 0 & 0 & 0 & 0 & 1 & 0 & 0\\ 0 & 0 & 0 & 0 & 0 & 0 & 0 & 0 & 0 & 0 & 0 & 0 & 0 & 0 & 0 & 0\\ 1 & 0 & 0 & 0 & 0 & 1 & 0 & 0 & 0 & 0 & 1 & 0 & 0 & 0 & 0 & 1\\ 0 & 0 & 0 & 0 & 0 & 0 & 0 & 0 & 0 & 0 & 0 & 0 & 0 & 0 & 0 & 0\\ 0 & 0 & 0 & 0 & 0 & 0 & 0 & 0 & 0 & 0 & 0 & 0 & 0 & 0 & 0 & 0\\ 0 & 0 & 1 & 0 & 0 & 0 & 0 & 1 & 1 & 0 & 0 & 0 & 0 & 1 & 0 & 0\\ 0 & 0 & 0 & 0 & 0 & 0 & 0 & 0 & 0 & 0 & 0 & 0 & 0 & 0 & 0 & 0\\ 1 & 0 & 0 & 0 & 0 & 1 & 0 & 0 & 0 & 0 & 1 & 0 & 0 & 0 & 0 & 1 \end{array}\right)
\]
Projecting from $(1, 0, 0, 0, 0, 0, 0, 0, 0, 0, 0, 0, 0, 0, 0, 0)^T$, be obtain
\[
H=E^{\bm{\tau}}\begin{pmatrix} 1 & 0 & 0 & 0 \\ 0 & 1 & 0 & 0  \\ 0 & 0 & 1 & 0 \\ 0 & 0 & 0 & 1 \end{pmatrix}
\]
where $E^{\bm{\tau}}$ is the material dependent parameter.  

\underline{Nearest neighbour interaction:} ${\rm\bf t}_\ell+\bm{\sigma}_\alpha= \bm{\tau}_b;\;{\rm\bf t}_{\ell^\prime}+\bm{\tau}_\beta=\bm{\tau}_a;\;\bm{\rho}_1= \bm{\tau}_a -\bm{\tau}_b $

The action of the coset representative on $\bm{\rho}_1$ produces 6 distinct vectors $\bm{\rho}_1, \ldots, \bm{\rho}_6$ or the corresponding phase factor $\exp[i{\rm\bf k}\cdot\bm{\rho}_r]$. The action of R permutes the 6 vectors or the corresponding phase factor $\exp[i{\rm\bf k}\cdot\bm{\rho}_r]$. Thus the representation matrices $D^{\bm{\rho}}_{FNN}(R)$ are $6\times 6$ matrices. We have $D^{\bm{\rho}}_{FNN}(E)=\mathbb{1}_{6\times 6}$. The generators are given by
\[
D^{\bm{\rho}}_{FNN}({\rm C}_6)=\begin{pmatrix}	
\bm{0} & \bm{0} & \sigma_1 \\
\sigma_1 & \bm{0} & \bm{0} \\
\bm{0} & \sigma_1 & \bm{0} 
\end{pmatrix}, \quad
D^{\bm{\rho}}_{FNN}({\rm I})=\begin{pmatrix}	
\sigma_1 & \bm{0} & \bm{0} \\
\bm{0} & \sigma_1 & \bm{0}\\
\bm{0} & \bm{0} & \sigma_1
\end{pmatrix}, \quad
D^{\bm{\rho}}_{FNN}({\rm C}_2^y)=\begin{pmatrix}
\sigma_0 & \bm{0} & \bm{0} \\
\bm{0} & \bm{0} & \sigma_0 \\
\bm{0} & \sigma_0 & \bm{0}
\end{pmatrix} 
\]
For the rest of the group element of ${\rm D}_{6h}$, $D^{\bm{\rho}}$ can be constructed from the generators. The projection operator $\mathcal{P}^{\Gamma_1}$ of dimension $24\times 24$ is constructed. Starting with  
\[(0, 0, 1, 0, 0, 0, 0, 0, 0, 0, 0, 0, 0, 0, 0, 0, 0, 0, 0, 0, 0, 0, 0, 0)^T\]
The solitary 1 correspond to $\exp\{i{\rm\bf k}\cdot\bm{\rho}_1\}$ and $H({\rm\bf k})_{ba}$ in the term specifying the partition. The following is obtained from projection:
\[(\underbrace{0, 0, 1, 0}_{e^{i{\rm\bf k}\cdot\bm{\rho}_1}}, \underbrace{0, 1, 0, 0}_{e^{i{\rm\bf k}\cdot\bm{\rho}_2}}, \underbrace{0, 0, 1, 0}_{e^{i{\rm\bf k}\cdot\bm{\rho}_3}}, \underbrace{0, 1, 0, 0}_{e^{i{\rm\bf k}\cdot\bm{\rho}_4}}, \underbrace{0, 0, 1, 0}_{e^{i{\rm\bf k}\cdot\bm{\rho}_5}}, \underbrace{0, 1, 0, 0}_{e^{i{\rm\bf k}\cdot\bm{\rho}_6}})^T\]
giving rise to a contribution to Hamiltonian from the nearest neighbour partition:
\[
H^{FNN}=t_{FNN}\left[\sum_{n=1}^3\exp(i{\rm\bf k}\cdot\bm{\rho}_{2n-1})\right]\begin{pmatrix}0 & 0\\  1 & 0 \end{pmatrix}+t_{FNN}\left[\sum_{n=1}^3\exp(i{\rm\bf k}\cdot\bm{\rho}_{2n})\right]\begin{pmatrix} 0 & 1 \\ 0 & 0\end{pmatrix}
\]
where $t_{FNN}$ is the material dependent parameter. For double group, the projection operator $\mathcal{P}^{\Gamma_1}$ is of dimension $96\times 96$. The contribution to the Hamiltonian from the first nearest neighbour partition is
\[
H^{FNN}=t_{FNN}\left[\sum_{n=1}^3\exp(i{\rm\bf k}\cdot\bm{\rho}_{2n-1})\right]\begin{pmatrix}\bm{0} & \bm{0}\\ \sigma_0 & \bm{0}\end{pmatrix}+t_{FNN}\left[\sum_{n=1}^3\exp(i{\rm\bf k}\cdot\bm{\rho}_{2n})\right]\begin{pmatrix}\bm{0} & \sigma_0\\ \bm{0} & \bm{0}\end{pmatrix}
\]

\underline{Second nearest neighbour interaction:} ${\rm\bf t}_\ell+\bm{\sigma}_\alpha= \bm{\tau}_a-\bm{a}_1;\;{\rm\bf t}_{\ell^\prime}+\bm{\tau}_\beta=\bm{\tau}_a;\;\;\bm{\rho}_1=\bm{\tau}_a-(\bm{\tau}_a-\bm{a}_1)= \bm{a}_1 $
There are 12 distinct $\bm{\rho}$ vectors. Whilst phase term from some of the vectors are numerically the same (e.g. $\bm{\rho}_1= \bm{\rho}_7$), they corresponds to hopping between different sites. The representation matrix for $D^{\bm{\rho}}_{SNN}(R)$ are
\[
D^{\bm{\rho}}_{SNN}({\rm C}_6)=\sigma_1\otimes\begin{pmatrix}	
\bm{0} & \bm{0} & \sigma_1 \\
\sigma_1 & \bm{0} & \bm{0} \\
\bm{0} & \sigma_1 & \bm{0} 
\end{pmatrix}, \quad
D^{\bm{\rho}}_{SNN}({\rm I})= \sigma_1\otimes\begin{pmatrix}	
\sigma_1 & \bm{0} & \bm{0} \\
\bm{0} & \sigma_1 & \bm{0}\\
\bm{0} & \bm{0} & \sigma_1
\end{pmatrix}, \quad
D^{\bm{\rho}}_{SNN}({\rm C}_2^y)=\sigma_0\otimes\begin{pmatrix}
\sigma_1 & \bm{0} & \bm{0} \\
\bm{0} & \bm{0} & \sigma_1 \\
\bm{0} & \sigma_1 & \bm{0}
\end{pmatrix}.
\]
Construct the projection operator $\mathcal{P}^{\Gamma_1}$ using the same method. Projecting from $(1, 0, 0, 0, \ldots)^T$ for single group and making use of the numerical equality between some of the radial vectors, one obtains
\[
H^{SNN}=t_{SNN}\left[\sum_{n=1}^6\exp(i{\rm\bf k}\cdot \bm{\rho}_n)\right]\begin{pmatrix} 1 & 0\\0 & 1\end{pmatrix}
\] 
For double group, two linearly independent terms can be projected out from the starting sub-space specified by ${\rm\bf t}_\ell+\bm{\sigma}_\alpha= \bm{\tau}_a-\bm{a}_1;\;{\rm\bf t}_{\ell^\prime}+\bm{\tau}_\beta=\bm{\tau}_a;\;\;\bm{\rho}_1=\bm{\tau}_a-(\bm{\tau}_a-\bm{a}_1)= \bm{a}_1 $. Projecting from $(1, 0, 0, 0 , 0, 1, 0, 0, \ldots)^T$ and $(1, 0, 0, 0 , 0, -1, 0, 0, \ldots)^T$, the contribution to the Hamiltonian from the second nearest neighbour partition is given by
\[
H^{SNN}=t_{SNN}\left[\sum_{n=1}^6\exp(i{\rm\bf k}\cdot \bm{\rho}_n)\right]\begin{pmatrix} 1 & 0 & 0 & 0\\ 0 & 1 & 0 & 0\\ 0 & 0 & 1 & 0 \\ 0 & 0 & 0 & 1\end{pmatrix}+t_{SNN}^\prime\left[\sum_{n=1}^6(-1)^{(n-1)}\exp(i{\rm\bf k}\cdot \bm{\rho}_n)\right]\begin{pmatrix} 1 & 0 & 0 & 0\\ 0 & -1 & 0 & 0\\ 0 & 0 & -1 & 0 \\ 0 & 0 & 0 & 1\end{pmatrix}
\]
Here $t_{SNN}$ and $t_{SNN}^\prime$ are two distinct material dependent parameters arising from two linearly independent vector transforming according to the trivial representation. The existence of the second symmetry permitted term is critical in the topology of the band structure of graphene. See Sec.\ref{sec:spinpz} for detail.

\bibliography{EBR2}{}
\bibliographystyle{apsrev4-1}
\end{document}